\documentclass[aps, pra, onecolumn, tightenlines, letterpaper, amsmath, amssymb, preprintnumbers, superscriptaddress, floatfix, longbibliography, nofootinbib]{revtex4-2}

\usepackage{dsfont}
\usepackage{amsmath}
\usepackage{amssymb}
\usepackage{physics}
\usepackage{tabu}
\usepackage{tabularx}
\usepackage{bm}
\usepackage{tikz}
\usetikzlibrary{quantikz}
\usepackage{textgreek}
\usepackage{multirow}
\usepackage{makecell}

\makeatletter
\newcommand{\fmarki}{*}
\newcommand{\fmarkii}{\ensuremath{\dagger}}
\newcommand{\fmarkiii}{\ensuremath{\ddagger}}
\newcommand{\fmarkiv}{\ensuremath{\mathsection}}
\newcommand{\fmarkv}{\ensuremath{\mathparagraph}}
\newcommand{\fmarkvi}{\ensuremath{\|}}
\newcommand{\nub}{\overline{\nu}}
\newcommand{\eb}{\overline{e}}
\newcommand{\ub}{\overline{u}}
\newcommand{\db}{\overline{d}}

\def\@fnsymbol#1{{\ifcase#1\or \fmarki\or \fmarkii\or \fmarkiii\or \fmarkiv\or \fmarkv\or \fmarkvi \else\@ctrerr\fi}}
\makeatother

\renewcommand{\fmarkvi}{\$}

\usepackage[hypertexnames=false]{hyperref}
\hypersetup{
    colorlinks=true,       
    linkcolor=blue,          
    citecolor=blue,        
    filecolor=blue,      
    urlcolor=blue           
}

\def\pslash{{ p\hskip-0.5em /}}

\newcolumntype{Y}{>{\centering\arraybackslash}X}

\makeatletter
\pretocmd\frontmatter@thefootnote{\color{black}}{}{}
\makeatother

\usepackage{orcidlink}

\begin{document}

\begin{figure}
  \vskip -1.cm
  \leftline{\includegraphics[width=0.15\textwidth]{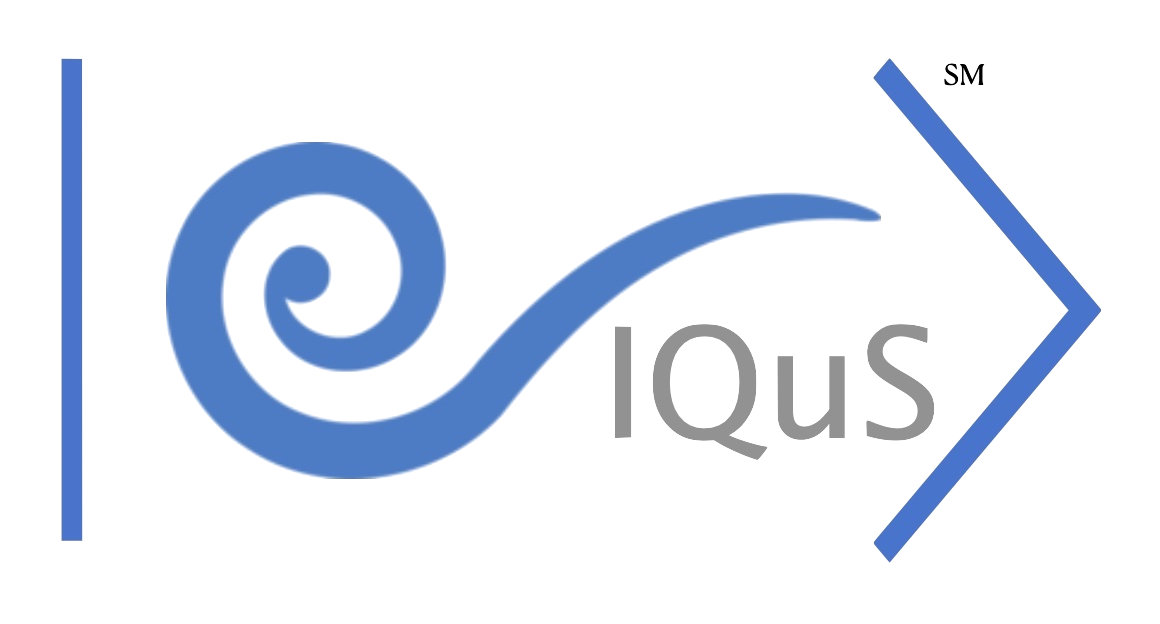}}
\end{figure}

\title{Preparations for Quantum Simulations of Quantum Chromodynamics in \texorpdfstring{\boldmath$1+1$}{1+1} Dimensions:
(II) Single-Baryon 
\texorpdfstring{$\beta$}{Beta}-Decay in Real Time}

\author{Roland C.~Farrell\,\orcidlink{0000-0001-7189-0424
}}
\email[Corresponding author, ]{rolanf2@uw.edu}
\affiliation{InQubator for Quantum Simulation (IQuS), Department of Physics, University of Washington, Seattle, WA 98195, USA.}
\author{Ivan A.~Chernyshev, \orcidlink{0000-0001-8289-1991}}
\email{ivanc3@uw.edu}
\affiliation{InQubator for Quantum Simulation (IQuS), Department of Physics, University of Washington, Seattle, WA 98195, USA.}
\author{Sarah J.~M.~Powell\,\orcidlink{0000-0002-5228-8291}}
\email{spow9@uw.edu}
\affiliation{Department of Physics and Astronomy, York University, Toronto, ON M3J 1P3, Canada.}
\author{Nikita A.~Zemlevskiy\,\orcidlink{0000-0002-0794-2389}}
\email{zemlni@uw.edu}
\affiliation{InQubator for Quantum Simulation (IQuS), Department of Physics, University of Washington, Seattle, WA 98195, USA.}
\author{Marc Illa\,\orcidlink{0000-0003-3570-2849}}
\email{marcilla@uw.edu}
\affiliation{InQubator for Quantum Simulation (IQuS), Department of Physics, University of Washington, Seattle, WA 98195, USA.}
\author{Martin J.~Savage\,\orcidlink{0000-0001-6502-7106}}
\email{mjs5@uw.edu}
\affiliation{InQubator for Quantum Simulation (IQuS), Department of Physics, University of Washington, Seattle, WA 98195, USA.}

\preprint{IQuS@UW-21-030, NT@UW-22-12}
\date{\today}

\begin{abstract}
\noindent
A framework for quantum simulations of
real-time weak decays of hadrons and nuclei 
in a 2-flavor lattice theory in one spatial dimension is
presented.
A single generation of the Standard Model is found to require 16 qubits per spatial lattice site after
mapping to spin operators via the Jordan-Wigner transformation.
Both quantum chromodynamics and flavor-changing weak interactions are included in the dynamics, 
the latter through four-Fermi effective operators. 
Quantum circuits that implement time evolution in this lattice theory are developed and  
run on Quantinuum's {\tt H1-1} 20-qubit trapped ion system to simulate the $\beta$-decay of a single baryon on one lattice site.
These simulations include the initial state preparation and are performed for both one and two Trotter time steps. 
The potential intrinsic error-correction properties of this type of lattice theory are discussed and the leading lattice Hamiltonian required to simulate $0\nu\beta\beta$-decay of nuclei induced by a neutrino Majorana mass term is provided.
\end{abstract}

\maketitle
\newpage{}
\section{Introduction}
\noindent
A quantitative exploration of hadronic decays and nuclear reaction dynamics resolved 
at very short time scales using quantum simulations will provide a new window into 
strong-interaction processes that lies beyond the capabilities of experiment.
In chemistry, the development of
femtosecond laser-pulse imaging in the 1980s~\cite{Zewail:1999}, allowed for reaction pathways to be studied in real time (for an overview, see Ref.~\cite{Durrani:2020}). 
Although a similar experimental procedure is not available for strong processes, it is expected that quantum simulations will provide analogous insight into hadronic dynamics.
Perhaps the simplest non-trivial class of such reactions to begin exploring is the
$\beta$-decay of low-lying hadrons and nuclei.
Single $\beta$-decay rates of nuclei have played a central role in defining the 
Standard Model (SM) of strong and electroweak processes~\cite{Glashow:1961tr,Higgs:1964pj,Weinberg:1967tq,Salam:1968rm}. They 
initially provided evidence that the weak (charged-current) 
quark eigenstates differ from the strong eigenstates, and, more recently, are
providing stringent tests of the unitarity of the Cabibbo-Kobayashi-Maskawa (CKM) matrix~\cite{Cabibbo:1963yz,Kobayashi:1973fv}.
For recent reviews of $\beta$-decay, see, e.g.,  Refs.~\cite{Gonz_lez_Alonso_2019,Hassan:2020hrj,Algora:2021,PhysRevC.102.045501}.
The four-Fermi operators responsible for $\beta$-decay~\cite{Feynman:1958ty} in the SM
emerge from operator production expansions (OPEs) 
of the non-local operators coming from the exchange of a charged-gauge boson ($W^-$) between quarks and leptons.
Of relevance to this work is the four-Fermi operator, which gives rise to the flavor changing quark process $d\rightarrow u e^-\overline{\nu}$.
In the absence of higher-order electroweak processes, including electromagnetism,
matrix elements of these operators factorize between the hadronic and leptonic sectors. This leaves, for example, a non-perturbative evaluation of $n\rightarrow p e^-\overline{\nu}$ for neutron decay, which is constrained significantly by the approximate global flavor symmetries of QCD.
Only recently have the observed systematics of $\beta$-decay rates of nuclei been understood without the need for phenomenological re-scalings of the axial coupling constant, 
$g_A$. 
As has long been anticipated, the correct decay rates are recovered when two-nucleon and higher-body interactions are included within the
effective field theories (EFTs) (or meson-exchange currents)~\cite{Baroni_2016,Krebs:2016rqz,Gysbers:2019uyb,Baroni_2021}. 
This was preceded by successes of EFTs in describing electroweak processes of few-nucleon systems through the inclusion of higher-body electroweak operators (not constrained by strong interactions alone), 
e.g., Refs.~\cite{Chen:1999tn,Butler:1999sv,Butler:2002cw,Baroni:2016xll,Li:2017udr,Baroni:2018fdn}.
The EFT framework describing nuclear $\beta$-decays involves contributions from ``potential-pion" and ``radiation-pion" exchanges~\cite{Kaplan:1998tg,Kaplan:1998we} 
(an artifact of a system of relativistic and non-relativistic particles~\cite{Grinstein:1997gv,Luke:1997ys})
and real-time simulations of these processes are expected to be able to isolate these distinct contributions.
Recently, 
the first Euclidean-space lattice QCD calculations of Gamow-Teller matrix elements in light nuclei 
(at unphysical light quark masses and without fully-quantified uncertainties) 
have been performed~\cite{Parreno:2021ovq}, 
finding results that are consistent with nature.

While $\beta$-decay is a well-studied and foundational area of sub-atomic physics, 
the double-$\beta$-decay of nuclei continues to present a theoretical challenge in the 
the search for physics beyond the SM.
For a recent review of the ``status and prospects" of $\beta\beta$-decay, see Ref.~\cite{Dolinski_2019}.
Although $2\nu\beta\beta$-decay is allowed in the SM, and is 
a second order $\beta$-decay process, 
$0\nu\beta\beta$-decay requires the violation of lepton number.
Strong interactions clearly play an essential role 
in the experimental detection of the $\beta\beta$-decay of nuclei, but
such contributions are non-perturbative and complex, and, for example, 
the EFT descriptions involve contributions from two- and higher-body correlated operators~\cite{Savage:1998yh,Shanahan:2017bgi,Tiburzi:2017iux,Cirigliano:2018hja,Cirigliano:2019vdj}.
The ability to study the real-time dynamics of such decay process in nuclei would likely 
provide valuable insight into the underlying strong-interaction mechanisms, and potentially offer first principles constraints beyond those from Euclidean-space lattice QCD.\footnote{
For discussions of the potential of lattice QCD to impact $\beta\beta$-decay, see, e.g., Refs.~\cite{Shanahan:2017bgi,Tiburzi:2017iux,Monge-Camacho:2019nby,Davoudi:2021noh,Cirigliano:2022oqy,USQCD:2022mmc,Detmold:2022jwu,Cirigliano:2022rmf}.}

Significant progress is being made toward quantum simulations of quantum field theories,
both conceptually and using NISQ quantum simulators and devices~\cite{Brower:1997ha,PhysRevA.73.022328,Zohar:2011cw,Zohar:2012ay,Tagliacozzo:2012vg,Banerjee:2012xg,Tagliacozzo:2012df,Zohar:2012xf,Zohar:2012ts,PhysRevLett.110.125303,Wiese:2013uua,Hauke:2013jga,Marcos:2014lda,Kuno:2014npa,Bazavov:2015kka,Kasper:2015cca,Brennen:2015pgn,Kuno:2016xbf,Martinez:2016yna,Zohar:2016iic,Kasper:2016mzj,Muschik:2016tws,Gonzalez-Cuadra:2017lvz,Banuls:2017ena,Dumitrescu:2018njn,Klco:2018kyo,Kaplan:2018vnj,Kokail:2018eiw,Lu:2018pjk,Yeter-Aydeniz:2018mix,Stryker:2018efp,Gustafson:2019mpk,Cloet:2019wre,Bauer:2019qxa,Shehab:2019gfn,Klco:2019xro,Alexandru:2019nsa,Davoudi:2019bhy,Avkhadiev:2019niu,Klco:2019evd,Magnifico:2019kyj,Gustafson:2021mky,Banuls:2019bmf,Klco:2019yrb,Mishra:2019xbh,Luo:2019vmi,Kharzeev:2020kgc,Mueller:2020vha,Shaw2020quantumalgorithms,PhysRevLett.122.050403,Yang_2020,Ji:2020kjk,Bender:2020jgr,Haase:2020kaj,Halimeh:2020ecg,Robaina:2020aqh,Yeter-Aydeniz:2020jte,Paulson:2020zjd,Halimeh:2020djb,VanDamme:2020rur,Barata:2020jtq,Milsted:2020jmf,Kasper:2020owz,Ott:2020ycj,Ciavarella:2021nmj,Bauer:2021gup,Gustafson:2021imb,ARahman:2021ktn,Atas:2021ext,Yeter-Aydeniz:2021olz,Davoudi:2021ney,Kan:2021nyu,Stryker:2021asy,Aidelsburger:2021mia,Zohar:2021nyc,Halimeh:2021vzf,Yeter-Aydeniz:2021mol,Knaute:2021xna,Wiese:2021djl,Meurice:2021pvj,Mueller:2021gxd,Riechert:2021ink,Halimeh:2021lnv,Zhang:2021bjq,Alam:2021uuq,Deliyannis:2021che,Perlin:2021xux,Funcke:2021aps,Gustafson:2021jtq,VanDamme:2021njp,Thompson:2021eze,Kan:2021blb,Ashkenazi:2021ieg,Alexandrou:2021ynh,Bauer:2021gek,Wang:2021iox,Iannelli:2021jhs,Ciavarella:2021lel,Nguyen:2021hyk,Yeter-Aydeniz:2022vuy,Hartung:2022hoz,Illa:2022jqb,Ji:2022qvr,Carena:2022kpg,Halimeh:2022rwu,Mildenberger:2022jqr,Deliyannis:2022uyh,Ciavarella:2022zhe,Caspar:2022llo,Bauer:2022hpo,Halimeh:2022pkw,Halimeh:2022mct,Raychowdhury:2022wbi,Dreher:2022scr,Rahman:2022rlg,Greenberg:2022kzy,Tuysuz:2022knj,Farrell:2022wyt,Atas:2022dqm,Bringewatt:2022zgq,Grabowska:2022uos,Asaduzzaman:2022bpi,Carena:2022hpz,Gustafson:2022xdt,Davoudi:2022uzo,Avkhadiev:2022ttx,Jang:2022nun}.
Simulations  of systems of quarks and gluons with 
$L=1,2$  spatial sites in 1-dimension, and one or two plaquettes 
of Yang-Mills gauge theories
are now being performed on quantum devices~\cite{Klco:2019evd,Atas:2021ext,Ciavarella:2021nmj,Ciavarella:2021lel,Illa:2022jqb,Rahman:2022rlg,Farrell:2022wyt,Atas:2022dqm}.
The spatial and temporal extent of such simulations are steadily increasing as both error-mitigation strategies and device performance improve. 
There has also been important algorithmic development on how decay widths and cross sections can be extracted from the computation of Green functions on quantum hardware~\cite{PhysRevD.102.094505}.
Recently, results of classical and quantum simulations of $SU(3)$ gauge theory with $N_f=1,2$ flavors of quarks in $1+1$ dimensions were presented~\cite{Farrell:2022wyt,Atas:2022dqm}.  
With a layout footprint of six qubits per flavor per lattice site using the Jordan-Wigner (JW) 
mapping~\cite{Jordan:1928wi}, one Trotter step of time evolution of one spatial site 
was simulated using IBM's superconducting quantum computers~\cite{IBMQ}, obtaining uncertainties at the percent level.

This paper extends our recent work~\cite{Farrell:2022wyt} 
to include flavor-changing 
weak interactions via a four-Fermi operator that generates the $\beta$-decay 
of hadrons and nuclei. The terms in the lattice Hamiltonian that generate a Majorana mass for the neutrinos are also given, although not included in the simulations.
Applying the JW mapping, it is found that a single generation of the SM (quarks and leptons) maps onto $16$ qubits per spatial lattice site. 
Using Quantinuum's {\tt H1-1} 20-qubit trapped ion quantum computer, the initial state of a baryon is both prepared and evolved with one and two Trotter steps on a single lattice site. 
Despite only employing a minimal amount of error mitigation, results at the 
$\sim 5\%$-level are obtained, consistent with the expectations.
Finally, we briefly comment on the potential of such hierarchical dynamics for error-correction purposes in quantum simulations.

\section{The \texorpdfstring{$\beta$}{Beta}-Decay Hamiltonian for Quantum Simulations in 1+1 Dimensions}
\noindent
In nature, the $\beta$-decays of neutrons and nuclei involve energy and momentum transfers related to the energy scales of nuclear forces and of isospin breaking. 
As these are much below the electroweak scale,
$\beta$-decay rates are well reproduced by matrix elements of 
four-Fermi effective interactions with $V-A$ structure~\cite{Feynman:1958ty,Sudarshan:1958vf}, of the form
\begin{equation}
    {\cal H}_\beta  = 
    \frac{G_F}{\sqrt{2}} \ V_{ud} \ 
    \overline{\psi}_u\gamma^\mu (1-\gamma_5)\psi_d\ 
    \overline{\psi}_e\gamma_\mu (1-\gamma_5)\psi_{\nu_e} 
    \ +\ {\rm h.c.}
    \ ,
    \label{eq:HbetaC}
\end{equation}
where $V_{ud}$ is the element of the CKM matrix for $d\rightarrow u$ transitions,
and $G_F$ is Fermi's coupling constant that is 
measured to be $G_F=1.1663787 (6) \times 10^{-5}~{\rm GeV}^{-2}$~\cite{Tiesinga:2021myr}.
This is the leading order (LO) SM result, obtained by matching amplitudes at 
tree-level, 
where $G_F/\sqrt{2} = g_2^2/(8 M_W^2)$ 
with $M_W$ the mass of the $W^\pm$ gauge boson
and $g_2$ the SU(2)$_L$ coupling constant.
Toward simulating the SM in $3+1$D, we consider
$1+1$D QCD containing $u$-quarks, $d$-quarks, electrons and electron neutrinos. 
For simplicity,
we model $\beta$-decay through a vector-like four-Fermi operator,
\begin{equation}
    {\cal H}_\beta^{1+1} =
    \frac{G}{\sqrt{2}} \
    \overline{\psi}_u\gamma^\mu \psi_d\ 
    \overline{\psi}_e\gamma_\mu \mathcal{C} \psi_{\nu}  
        \ +\ {\rm h.c.}
    \ ,
    \label{eq:HbetaC1}
\end{equation}
where $\mathcal{C} = \gamma_1$ is the charge-conjugation operator 
whose purpose will become clear. 
Appendices~\ref{app:betaSM} and~\ref{app:beta1p1} provide details on
calculating the single-baryon $\beta$-decay rates in the
infinite volume and continuum limits in the SM and in the $1+1$D model considered here.

The strong and weak interactions can be mapped
onto the finite-dimensional Hilbert space provided by a quantum computer
by using the Kogut-Susskind (KS) Hamiltonian formulation of
lattice gauge theory~\cite{Kogut:1974ag,Banks:1975gq}. 
The KS discretization of the fields is such that
$L$ spatial lattice sites 
are split into  $2L$ fermion sites 
that separately accommodate 
fermions (even sites) and anti-fermions (odd sites).
For the $\beta$-decay of baryons, the strong and the weak KS Hamiltonian (in axial gauge) 
has the form~\cite{Banuls:2017ena,Atas:2021ext,Farrell:2022wyt,Atas:2022dqm}
\begin{equation}
    H = H_{{\rm quarks}} + H_{{\rm leptons}} +  H_{{\rm glue}} + H_{\beta}
    \ ,
\end{equation}
where
\begin{align}
    H_{\rm{quarks}} 
    =&\ 
    \sum_{f=u,d}\left[
        \frac{1}{2 a} \sum_{n=0}^{2L-2} \left ( \phi_n^{(f)\dagger} \phi_{n+1}^{(f)}
        \ +\ {\rm h.c.} \right ) 
    \: + \: 
    m_f \sum_{n=0}^{2L-1} (-1)^{n} \phi_n^{(f)\dagger} \phi_n^{(f)} 
    \right] \ ,
    \nonumber\\
    H_{\rm{leptons}} 
    =&\ 
        \sum_{f=e,\nu}\left[
    \frac{1}{2 a} \sum_{n=0}^{2L-2} \left ( \chi_n^{(f)\dagger} \chi_{n+1}^{(f)}
        \ +\ {\rm h.c.} \right ) 
    \: + \: 
    m_f \sum_{n=0}^{2L-1} (-1)^{n} \chi_n^{(f)\dagger} \chi_n^{(f)} 
\right]  \ ,
    \nonumber\\
    H_{{\rm glue}}
    =&\ 
    \frac{a g^2}{2} 
    \sum_{n=0}^{2L-2} 
    \sum_{a=1}^8
    \left ( \sum_{m\leq n} Q^{(a)}_m \right )^2 \ ,
    \nonumber\\
    H_{{\rm \beta}}
    =&\
    \frac{G}{a \sqrt{2}} 
    \sum_{l=0}^{L-1} \bigg [
    \left (\phi_{2l}^{(u)\dagger} \phi_{2l}^{(d)} + \phi_{2l+1}^{(u)\dagger} \phi_{2l+1}^{(d)} \right ) \left (\chi_{2l}^{(e)\dagger} \chi_{2l+1}^{(\nu)} - \chi_{2l+1}^{(e)\dagger} \chi_{2l}^{(\nu)}\right ) 
    \nonumber\\
    &+
    \left ( \phi_{2l}^{(u)\dagger} \phi_{2l+1}^{(d)} + \phi_{2l+1}^{(u)\dagger} \phi_{2l}^{(d)} \right )
    \left (\chi_{2l}^{(e)\dagger} \chi_{2l}^{(\nu)} - \chi_{2l+1}^{(e)\dagger} \chi_{2l+1}^{(\nu)}\right )+
    {\rm h.c.} \bigg ] \ .
    \label{eq:KSHam}
\end{align}
The masses of the $u$-, $d$-quarks, electron and neutrino (Dirac) are $m_{u,d,e,\nu}$,
and the strong and weak coupling constants are $g$ and $G$. The $SU(3)$ charges are
$Q_m^{(a)}$, and
$\phi^{(u,d)}_n$ are the $u$- and $d$-quark field operators (which both transform in the fundamental representation of $SU(3)$, and hence the sum over color indices has been suppressed). The electron and  neutrino field operators are
$\chi^{(e,\nu)}_n$, and for the remainder of this paper the lattice spacing, $a$, will be set to unity.
We emphasize that the absence of gluon fields is due to the choice of axial gauge, whereas the lack of weak gauge fields is due to the
consideration of a low energy effective theory in which the heavy weak gauge bosons have been integrated out. 
This results in, for example, the absence of parallel transporters in the fermion kinetic terms.
\begin{figure}[!t]
    \centering
    \includegraphics[width=15cm]{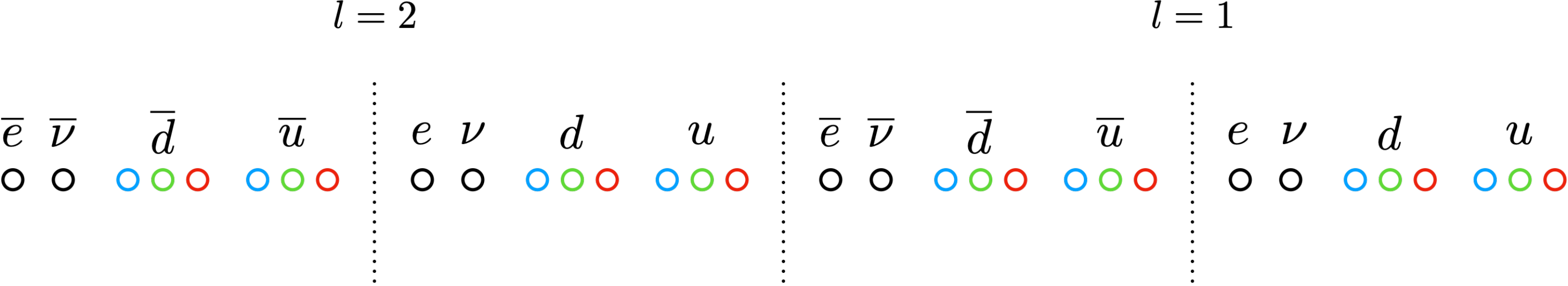}
    \caption{
    The qubit layout of a $L=2$ lattice,
    where fermions and anti-fermions are grouped together (which will be preferred if electromagnetism is included).  This layout extends straightforwardly to $L>2$.
}
    \label{fig:L1layout}
\end{figure}

The JW mapping of the Hamiltonian in Eq.~(\ref{eq:KSHam}) to qubits, 
arranged as shown in Fig.~\ref{fig:L1layout},
is given by 
\begin{align}
    H_{\rm{quarks}} 
     \rightarrow &\
    \frac{1}{2} \sum_{l=0}^{L-1} \sum_{f=u,d}\sum_{c=0}^{2}  m_f\left ( Z_{l,f,c} - Z_{l,\overline{f},c} + 2\right )  \nonumber \\
    & -\frac{1}{2} \sum_{l=0}^{L-1} \sum_{f=u,d} \sum_{c=0}^{2}\left [ \sigma^+_{l,f,c} Z^7 \sigma^-_{l,\overline{f},c}  + (1-\delta_{l,L-1})  \sigma^+_{l,\overline{f},c} Z^7 \sigma^-_{l+1,f,c} + {\rm h.c.} \right ]\ , \nonumber \\[4pt]
    H_{\rm{leptons}} 
    \rightarrow  &\ \frac{1}{2} \sum_{l=0}^{L-1} \sum_{f=e,\nu}  m_f\left ( Z_{l,f} - Z_{l,\overline{f}} + 2\right )  \: - \: \frac{1}{2} \sum_{l=0}^{L-1} \sum_{f=e,\nu} \left [ \sigma^+_{l,f} Z^7 \sigma^-_{l,\overline{f}}  + (1-\delta_{l,L-1})  \sigma^+_{l,\overline{f}} Z^7 \sigma^-_{l+1,f} + {\rm h.c.} \right ] \ ,
    \nonumber\\[4pt]
    H_{{\rm glue}}
    \rightarrow  &\ \frac{g^2}{2} \sum_{n=0}^{2L-2}(2L-1-n)\left( \sum_{f=u,d} Q_{n,f}^{(a)} \, Q_{n,f}^{(a)} \ + \
        2 Q_{n,u}^{(a)} \, Q_{n,d}^{(a)}
         \right)    \nonumber \\
         &+ \: g^2 \sum_{n=0}^{2L-3} \sum_{m=n+1}^{2L-2}(2L-1-m) \sum_{f=u,d} \sum_{f'=u,d} Q_{n,f}^{(a)} \, Q_{m,f'}^{(a)} \ ,
    \nonumber\\[4pt]
    H_{{\rm \beta}}
    \rightarrow  &\
    \frac{G}{\sqrt{2}}\sum_{l = 0}^{L-1}\sum_{c=0}^2\bigg ( \sigma^-_{l,\nub} Z^6 \sigma^+_{l,e} \sigma^-_{l,d,c} Z^2 \sigma^+_{l,u,c} \: - \: \sigma^+_{l,\eb} Z^8 \sigma_{l,\nu}^- \sigma_{l,d,c}^- Z^2 \sigma^+_{l,u,c} 
    \: - \: \sigma^-_{l,\nub} Z^{2-c} \sigma^-_{l,\db,c} \sigma^+_{l,\ub,c} Z^c \sigma^+_{l,e} \nonumber \\
    & + \: \sigma^+_{l,\eb}Z^{3-c}\sigma^-_{l,\db,c} \sigma^+_{l,\ub,c} Z^{1+c} \sigma^-_{l,\nu} \: - \: \sigma^-_{l,\db,c} Z^{3+c} \sigma^+_{l,e} \sigma^-_{l,\nu} Z^{5-c} \sigma^+_{l,u,c} \: - \: \sigma^+_{l,\eb} \sigma^-_{l,\nub} \sigma^-_{l,\db,c}Z^{10}\sigma^+_{l,u,c} \nonumber \\ 
    & - \: \sigma^+_{l,\ub,c} Z^c \sigma^+_{l,e} \sigma^-_{l,\nu} Z^{2-c} \sigma^-_{l,d,c} \: - \: \sigma^+_{l,\eb} \sigma^-_{l,\nub} \sigma^+_{l,\ub,c} Z^4 \sigma^-_{l,d,c} \: + \: {\rm h.c.}\bigg )
     \ ,
\label{eq:KSHamJWmap}
\end{align}
where the sums of products of color charges are
\begin{align}
    Q_{n,f}^{(a)} \, Q_{n,f}^{(a)} = & \ 1 - \frac{1}{3}\sum_{c=0}^1 \sum_{c' >c}Z_{n,f,c} Z_{n,f,c'} \ ,  \nonumber \\[4pt]
    Q_{n,f}^{(a)} \, Q_{m,f'}^{(a)} = & \ \frac{1}{2}\sum_{c=0}^{1} \sum_{c'>c} \left( \sigma^+_{n,f,c} Z^{c'-c-1} \sigma^-_{n,f,c'} \sigma^-_{m,f',c} Z^{c'-c-1} \sigma^+_{m,f',c'} + {\rm h.c.} \right )   + \frac{1}{24}\sum_{c=0}^{2} \sum_{c'=0}^2( 3 \delta_{c c'} - 1 ) Z_{n,f,c}Z_{m,f',c'} \ ,
    \label{eq:QnfQmfp}
\end{align}
and the repeated $SU(3)$ adjoint indices, $a=1,2,...,8$, are summed over. 
The index $l$ labels the spatial lattice site, $f \ (\overline{f})$ labels the (anti)fermion flavor and $c=0,1,2$ corresponds to red, green and blue colors.  
In the staggered mapping, there are gauge-field links every half of a spatial site and, as a result, the color charges are labelled by a half site index, $n$.
The spin raising and lowering operators are $\sigma^{\pm} = \frac{1}{2}(\sigma^x \pm i \sigma^y)$, $Z=\sigma^z$ and
unlabelled $Z$s act on the sites between the $\sigma^{\pm}$, e.g., $\sigma^-_{l,d,r} Z^2 \sigma^+_{l,u,r} = \sigma^-_{l,d,r} Z_{l,u,b} Z_{l,u,g} \sigma^+_{l,u,r}$. 
Constants have been added to the mass terms to ensure that all basis
states contribute positive mass.

\subsection{Efficiently Mapping the \texorpdfstring{$L=1$}{L=1} Hamiltonian to Qubits}
\begin{figure}[!t]
    \centering
    \includegraphics[width=15cm]{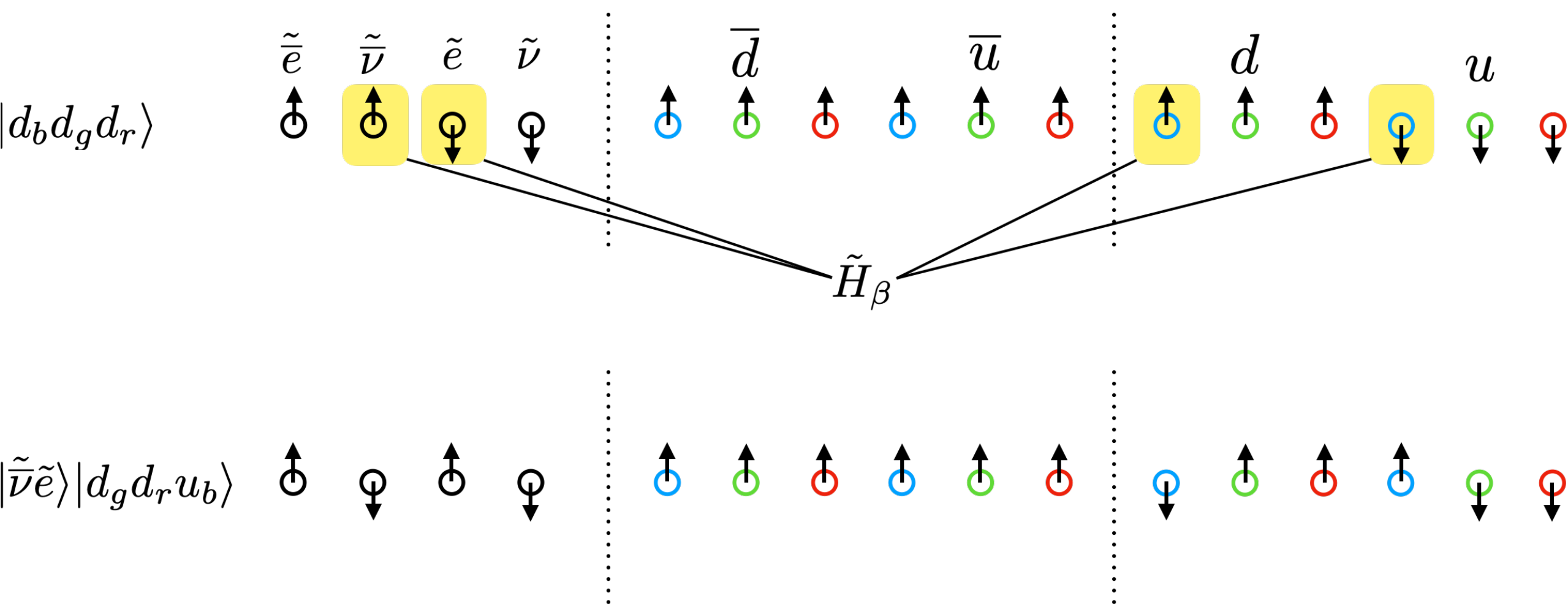}
    \caption{
    The $L=1$ lattice qubit layout of one generation of the SM that is used in this paper for quantum simulation. 
    Fermion (anti-fermion) sites are occupied when the spin is up (down), and the spins at the lepton sites represent occupation in the tilde basis.
    Specifically, 
    the example of $\ket{d_b d_g d_r}$ (upper lattice) decaying to $\ket{ \tilde{\overline{\nu}} \tilde{e}}\ket{d_g d_r u_b}$ (lower lattice) through one application of $\tilde{H}_{\beta}$ 
    in Eq.~(\ref{eq:tildeBeta}) is shown.
    }
    \label{fig:EWlayoutTilde}
\end{figure}

\noindent
To accommodate the capabilities of current devices, 
the quantum simulations performed in this work involve only a single spatial site, $L=1$, 
where the structure of the Hamiltonian can be simplified.
In particular, without interactions between leptons, it is convenient to work with field operators that create and annihilate eigenstates of the free lepton Hamiltonian, $H_{{\rm leptons}}$.
These are denoted by  ``tilde operators"~\cite{Farrell:2022wyt},
which create the open-boundary-condition (OBC) analogs of plane waves. 
In the tilde basis with the JW mapping, the lepton Hamiltonian is diagonal and becomes
\begin{equation}
    \tilde{H}_{{\rm leptons}} = \lambda_{\nu}(\tilde{\chi}^{(\nu) \dagger}_0 \tilde{\chi}^{(\nu)}_0-\tilde{\chi}^{(\nu) \dagger}_1 \tilde{\chi}^{(\nu)}_1)  + \lambda_{e}(\tilde{\chi}^{(e) \dagger}_0 \tilde{\chi}^{(e)}_0-\tilde{\chi}^{(e) \dagger}_1 \tilde{\chi}^{(e)}_1) \ \rightarrow \ \frac{\lambda_{\nu}}{2}(Z_{\nu} - Z_{\overline{\nu}})  + \frac{\lambda_{e}}{2}(Z_{e} - Z_{\overline{e}}) 
    \ ,
    \label{eq:tildeLep}
\end{equation}
where $\lambda_{\nu,e} = \frac{1}{2}\sqrt{1+4m_{\nu,e}^2}$. 
The $\beta$-decay operator in Eq.~(\ref{eq:KSHam}) becomes
\begin{alignat}{2}
    \tilde{H}_{\beta} = \frac{G}{\sqrt{2}}\Bigg \{ \left ( \phi_0^{(u)\dagger} \phi_{0}^{(d)}  +  \phi_{1}^{(u)\dagger} \phi_{1}^{(d)} \right ) \bigg [&\frac{1}{2}(s_+^e s_-^{\nu} \ - \ s_-^e s_+^{\nu})\left (\tilde{\chi}_0^{(e)\dagger} \tilde{\chi}_{0}^{(\nu)}  + \tilde{\chi}_1^{(e)\dagger} \tilde{\chi}_1^{(\nu)}\right ) \nonumber \\
     + \ &\frac{1}{2}( s_+^e s_+^{\nu} \ + \ s_-^e s_-^{\nu})\left (\tilde{\chi}_0^{(e)\dagger} \tilde{\chi}_{1}^{(\nu)} - \tilde{\chi}_1^{(e)\dagger} \tilde{\chi}_0^{(\nu)}\right )\bigg] \nonumber \\
    + \ \left ( \phi_0^{(u)\dagger} \phi_{1}^{(d)}  + \phi_{1}^{(u)\dagger} \phi_{0}^{(d)} \right )\bigg [
    &\frac{1}{2}(s_+^e s_+^{\nu} \ -\ s_-^e s_-^{\nu})\left (\tilde{\chi}_0^{(e)\dagger} \tilde{\chi}_{0}^{(\nu)}  - \tilde{\chi}_1^{(e)\dagger} \tilde{\chi}_1^{(\nu)}\right ) \nonumber \\ 
    - \ &\frac{1}{2}(s_+^e s_-^{\nu} \ +\  s_-^e s_+^{\nu})\left (\tilde{\chi}_0^{(e)\dagger} \tilde{\chi}_{1}^{(\nu)} +  \tilde{\chi}_1^{(e)\dagger} \tilde{\chi}_0^{(\nu)}\right ) \bigg ] \ + \ {\rm h.c.} \Bigg \}  \ ,
\label{eq:HbetaTilNoJW}
\end{alignat}
where $s^{\nu,e}_\pm = \sqrt{1\pm m_{\nu,e}/\lambda_{\nu,e}}$. 
In our simulations, the initial state of the quark-lepton 
system is prepared in a strong eigenstate with baryon number $B=+1$ in the quark sector 
and the vacuum, $\lvert \Omega \rangle_{{\rm lepton}}$, in the lepton sector.
One of the benefits of working in the tilde basis is that the vacuum satisfies $\tilde{\chi}^{(e,v)}_0\lvert \Omega \rangle_{{\rm lepton}} = \tilde{\chi}^{(e,v) \dagger}_1 \lvert \Omega \rangle_{{\rm lepton}} = 0$, and the terms in the first and third lines of Eq.~(\ref{eq:HbetaTilNoJW}) do not contribute to $\beta$-decay. For the processes we are interested in, this results in an effective $\beta$-decay operator of the form
\begin{alignat}{2}
    \tilde{H}_{\beta} = \frac{G}{\sqrt{2}}\Bigg \{ \left ( \phi_0^{(u)\dagger} \phi_{0}^{(d)}  +  \phi_{1}^{(u)\dagger} \phi_{1}^{(d)} \right ) \bigg [&\frac{1}{2}( s_+^e s_+^{\nu} \ + \ s_-^e s_-^{\nu})\left (\tilde{\chi}_0^{(e)\dagger} \tilde{\chi}_{1}^{(\nu)}  - \tilde{\chi}_1^{(e)\dagger} \tilde{\chi}_0^{(\nu)}\right )\bigg] \nonumber \\
    - \ \left ( \phi_0^{(u)\dagger} \phi_{1}^{(d)} + \phi_{1}^{(u)\dagger} \phi_{0}^{(d)} \right )\bigg [  &\frac{1}{2}(s_+^e s_-^{\nu} \ +\  s_-^e s_+^{\nu})\left (\tilde{\chi}_0^{(e)\dagger} \tilde{\chi}_{1}^{(\nu)} +  \tilde{\chi}_1^{(e)\dagger} \tilde{\chi}_0^{(\nu)}\right ) \bigg ] \ + \ {\rm h.c.} \Bigg \}  \ .
\label{eq:HbetaTilNoVac}
\end{alignat}
The insertion of 
the charge-conjugation matrix,
$\mathcal{C}$, in the continuum operator, Eq.~(\ref{eq:HbetaC1}),
is
necessary to obtain a $\beta$-decay operator that does not annihilate the lepton vacuum.
To minimize the length of the string of $Z$s in the JW mapping, the lattice layout in Fig.~\ref{fig:EWlayoutTilde} is used.
In this layout, the hopping piece of $H_{{\rm quarks}}$ has only 5 $Z$s between the quark and antiquark raising and lowering operators and the $\beta$-decay operator is
\begin{align}
\!\!\!\!\!\! \tilde{H}_{\beta }
    \rightarrow
    \frac{G}{\sqrt{2}}
      \sum_{c=r,g,b} \bigg [&\frac{1}{2}( s_+^e s_+^{\nu} \ + \ s_-^e s_-^{\nu})\left ( \sigma^-_{\overline{\nu}} \sigma^+_e  \ - \  \sigma^+_{\overline{e}} Z^2 \sigma^-_{\nu} \right )\left (\sigma^-_{d,c}Z^2 \sigma^+_{u,c} \: + \: \sigma^-_{\overline{d},c} Z^2 \sigma^+_{\overline{u},c}\right ) \nonumber \\
     - \ &\frac{1}{2}(s_+^e s_-^{\nu} \ +\  s_-^e s_+^{\nu})\left (\sigma^-_{\overline{\nu}} \sigma^+_e \ + \  \sigma^+_{\overline{e}} Z^2 \sigma^-_{\nu} \right )\left ( \sigma^-_{\overline{d},c} Z^8 \sigma^+_{u,c} \: + \: \sigma^+_{\overline{u},c} Z^2 \sigma^-_{d,c} \right ) \ + \  {\rm h.c.} \bigg ]  \ .
     \label{eq:tildeBeta}
\end{align}
In total, the $L=1$ system requires $16$ ($12$ quark and $4$ lepton) 
qubits. See App.~\ref{app:fullHam} for the complete $L=1$ Hamiltonian in terms of qubits.

\subsection{A Majorana Mass for the Neutrino}
\noindent
Although not relevant to the simulations performed in Sec.~\ref{sec:BetaSim}, it is of current interest to consider the inclusion of a Majorana mass term for the neutrinos.
A Majorana mass requires and induces the violation of lepton number by
$|\Delta L| = 2$, and is not present in the minimal SM, defined by dim-4 operators.
However, the Weinberg operator~\cite{Weinberg:1979sa} enters at dim-5 and generates an effective Majorana mass for the neutrinos,
\begin{align}
{\cal L}^{{\rm Weinberg}} 
&= \frac{1}{ 2\Lambda} 
\left( \overline{L}^c \epsilon \phi \right)
\left(\phi^T \epsilon L \right)
    \ +\ {\rm h.c.}
\ ,\ \ 
L = \left( \nu , e \right)^T_L
\ ,\ \ 
\phi = \left( \phi^+ , \phi^0 \right)^T
\ ,\ \ 
\langle \phi \rangle = \left( 0 , v/\sqrt{2} \right)^T
\ ,\ \ \epsilon = i\sigma_2
\ \ ,
\nonumber\\
&\rightarrow
-\frac{v^2}{4\Lambda}  \overline{\nu}^c_L \nu_L
    \ +\ {\rm h.c.}
\ +\ .... 
\label{eq:maja}
\end{align}
where $\phi$ is the Higgs doublet, 
$L^c$ denotes the charge-conjugated left-handed lepton doublet,
$v$ is the Higgs vacuum expectation value and $\Lambda$ is a high energy scale characterizing physics beyond the SM. 
The ellipsis denote interaction terms involving components of the Higgs doublet fields and the leptons.
This is the leading contribution beyond the minimal SM, 
but does not preclude contributions from other sources.
On a $1+1$D lattice there is only a single 
$\lvert \Delta L \rvert = 2$ local operator 
with the structure of a mass term
and, using the JW mapping along with the qubit layout in Fig.~\ref{fig:L1layout}, is of the form
\begin{equation}
H_{\rm Majorana} =
\frac{1}{2} m_M 
     \sum_{n={\rm even}}^{2L-2} 
\left( 
\chi_n^{(\nu)}
\chi_{n+1}^{(\nu)} \:
+ \: {\rm h.c.}
\right)
 \ \rightarrow
 \frac{1}{2} m_M  \sum_{l = 0}^{L-1} 
\left( 
\sigma^+_{l,\nu}\ 
Z^7
\sigma^+_{l,\nub} \: + \: {\rm h.c.}
\right)
\ .
\label{eq:majaHami}
\end{equation}
While the operator has support on a single spatial lattice site, it does not contribute to 
$0\nu\beta \beta$-decay on a lattice with only a single spatial site.   
This is because the processes that it could potentially induce, such as
$\Delta^-\Delta^-\rightarrow \Delta^0\Delta^0 e^- e^-$,
are Pauli-blocked by the single electron site.  
At least two spatial sites are required for any such process producing two electrons in the final state.

\section{Quantum Simulations of the \texorpdfstring{$\beta$}{Beta}-Decay of One Baryon on One Lattice Site}
\label{sec:BetaSim}
\noindent
In this section, quantum simulations of the $\beta$-decay of a single baryon are performed
in $N_f=2$ flavor QCD with $L=1$ spatial lattice site.
The required quantum circuits to perform one and two Trotter steps of time evolution were developed and run on the Quantinuum {\tt H1-1} $20$ qubit trapped ion quantum computer and its simulator {\tt H1-1E}~\cite{QuantHoney,h1-1e}.

\subsection{Preparing to Simulate \texorpdfstring{$\beta$}{Beta}-Decay}
\label{sec:BetaSimA}
\noindent
It is well known that, because of confinement, the  energy eigenstates (asymptotic states) of QCD 
are color-singlet hadrons, which are composite objects of quarks and gluons.
On the other hand, 
the operators responsible for $\beta$-decay, given in Eq.~(\ref{eq:tildeBeta}), 
generate transitions between $d$- and $u$-quarks.
As a result,  observable effects of $\tilde{H}_{\beta}$, in part, 
are found in transitions between 
hadronic states whose matrix elements depend on the distribution of the quarks within. 
Toward quantum simulations of the $\beta$-decay of neutrons and nuclei more generally, 
the present work focuses on the decay of a single baryon.

Generically, three elements are required for real-time quantum simulations of 
the $\beta$-decay of baryons:
\begin{enumerate}
    \item 
    Prepare the initial hadronic state that will subsequently undergo $\beta$-decay. 
    In this work, this is one of the single-baryon states (appropriately selected in the spectrum) 
    that is an eigenstate of the strong Hamiltonian alone, 
    i.e., the weak coupling constant is set equal to $G=0$.
    \item 
    Perform (Trotterized) time-evolution using the full ($G\neq0$) Hamiltonian.
    \item 
    Measure one or more of the lepton qubits. 
    If leptons are detected, then $\beta$-decay has occurred.
\end{enumerate}
In $1+1$D, Fermi statistics 
preclude the existence of a light isospin $I=1/2$ nucleon,
and the lightest baryons are in an $I=3/2$ multiplet 
$(\Delta^{++}, \Delta^+, \Delta^0, \Delta^-)$
(using the standard electric charge assignments of the up and down quarks). 
We have  chosen to simulate the decay 
$\Delta^- \to \Delta^0 + e + \overline{\nu}$, which, at the quark level, involves 
baryon-interpolating operators with the quantum numbers of
$ddd\rightarrow udd$.

In order for $\beta$-decay to be kinematically allowed, 
the input-parameters of the theory must be such that 
$M_{\Delta^-} > M_{\Delta^0} + M_{\overline{\nu}} + M_{e}$. 
This is accomplished through tuning the parameters of the Hamiltonian.
The degeneracy in the iso-multiplet is lifted 
by using different values for the  up and down quark masses. 
It is found that the choice of parameters, $m_u=0.9$, $m_d=2.1$, $g=2$ and $m_{e,\nu} = 0$ 
results in the desired hierarchy of baryon and lepton masses. 
The relevant part of the spectrum, obtained from an exact diagonalization of the Hamiltonian, 
is shown in Table~\ref{tab:betaMass}. Although kinematically allowed, multiple instances of $\beta$-decay cannot occur for $L=1$ as there can be at most one of each (anti)lepton.
\begin{table}[!ht]
\renewcommand{\arraystretch}{1.2}
\begin{tabularx}{0.5\textwidth}{||c | Y ||} 
\hline
\multicolumn{2}{||c||}{Energy of states relevant for $\beta$-decay (above the vacuum)} \\
 \hline
 State & Energy Gap\\
 \hline\hline
 $\Delta^{++}$ & 2.868 \\ 
 \hline
 $\Delta^{++}$ + $2 l$ & 3.868\\
 \hline
 $\Delta^+$ & 4.048 \\
 \hline
 $\Delta^{++}$ + $4l$  & 4.868\\
 \hline
 $\Delta^{+}$ + $2l$  & 5.048\\
 \hline
 $\Delta^0$ & 5.229 \\
 \hline
 $\Delta^{+}$ + $4l$& 6.048\\
 \hline
 $\Delta^0$ + $2l$ & 6.229 \\
 \hline
 $\Delta^-$ & 6.409 \\
 \hline
\end{tabularx}
\caption{
The energy gap above the vacuum of states relevant for $\beta$-decays of single baryons 
with $m_u = 0.9$, $m_d = 2.1$, $g=2$ and $m_{e,\nu} = 0$. The leptons are degenerate in energy and collectively denoted by $l$.}
\renewcommand{\arraystretch}{1}
\label{tab:betaMass}
\end{table}
Note that even though $m_{e,\nu} = 0$, 
the electron and neutrino are gapped due to the finite spatial volume.

\begin{figure}[!ht]
    \centering
    \includegraphics[width=\textwidth]{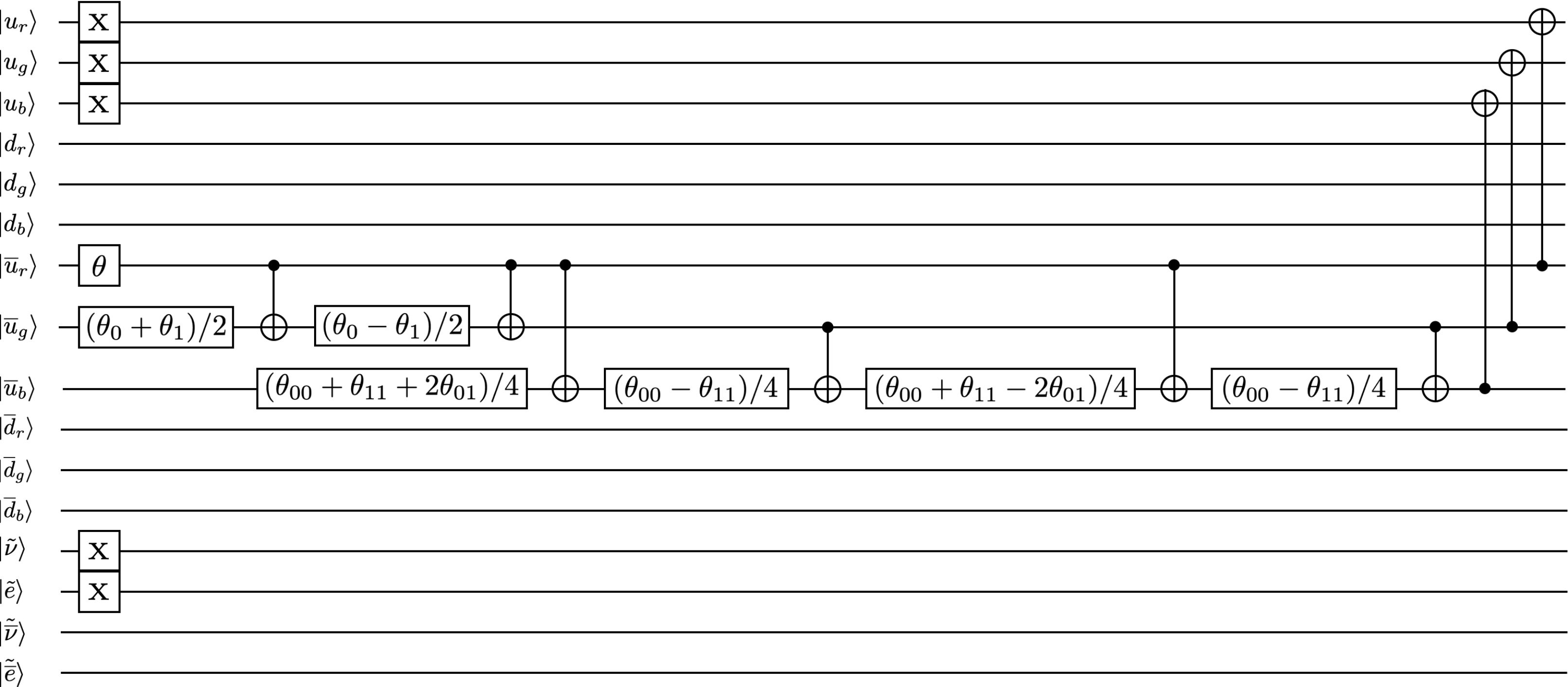}
    \caption{
    A quantum circuit that can be used to prepare the $\Delta^-$-baryon on $L=1$ spatial site.}
    \label{fig:DMVQE}
\end{figure}

To prepare the $\Delta^-$ initial state, 
we exploit the observation made in Ref.~\cite{Farrell:2022wyt} 
that the stretched-isospin eigenstates of the $\Delta$-baryons, 
with third component of isospin $I_3 = \pm 3/2$,
factorize between the $u$ and $d$ flavor sectors for $L=1$. 
Therefore, the previously developed 
Variational Quantum Eigensolver (VQE)~\cite{Peruzzo_2014} 
circuit~\cite{Farrell:2022wyt} used to prepare the one-flavor vacuum can be used to initialize the two-flavor $\Delta^-$ wave function.
This is done by initializing the vacuum in the lepton sector, 
preparing the state $\ket{d_r d_g d_b}$ in the $d$-sector, 
and applying the VQE circuit to produce the $u$-sector vacuum. 
In the tilde basis, the lepton vacuum is the unoccupied state (trivial vacuum), 
and the complete state-preparation circuit is shown in Fig.~\ref{fig:DMVQE}, 
where $\theta$ is shorthand for RY($\theta$). 
The rotation angles are related by
\begin{equation}
    \theta_0 = -2 \sin^{-1} \left[ \tan(\theta/2) \, \cos(\theta_{1}/2)  \right] \ \: ,  \ \: \theta_{00} = -2 \sin^{-1}\left[ \tan(\theta_{0}/2) \, \cos(\theta_{01}/2) \right] \ \: , \ \:
    \theta_{01} = -2 \sin^{-1}\left[ \cos(\theta_{11}/2)\, \tan(\theta_{1}/2) \right]
    \label{eq:angleconst}
\end{equation}
and, for $m_u = 0.9$ and $g=2$,\footnote{
The $\overline{u}$ and $u$ parts of the lattice are separated by a fully packed $d$ sector which implies that the part of the wavefunctions with odd numbers of anti-up quarks have relative minus signs compared to the one-flavor vacuum wavefunction.
}
\begin{equation}
    \theta = 0.2256 \ \ , \ \ \theta_1 = 0.4794 \ \ , \ \ \theta_{11} = 0.3265 \ .
\end{equation}
In total, state preparation requires the application of $9$ CNOT gates.

\begin{figure}[!ht]
    \centering
    \includegraphics[width=12cm]{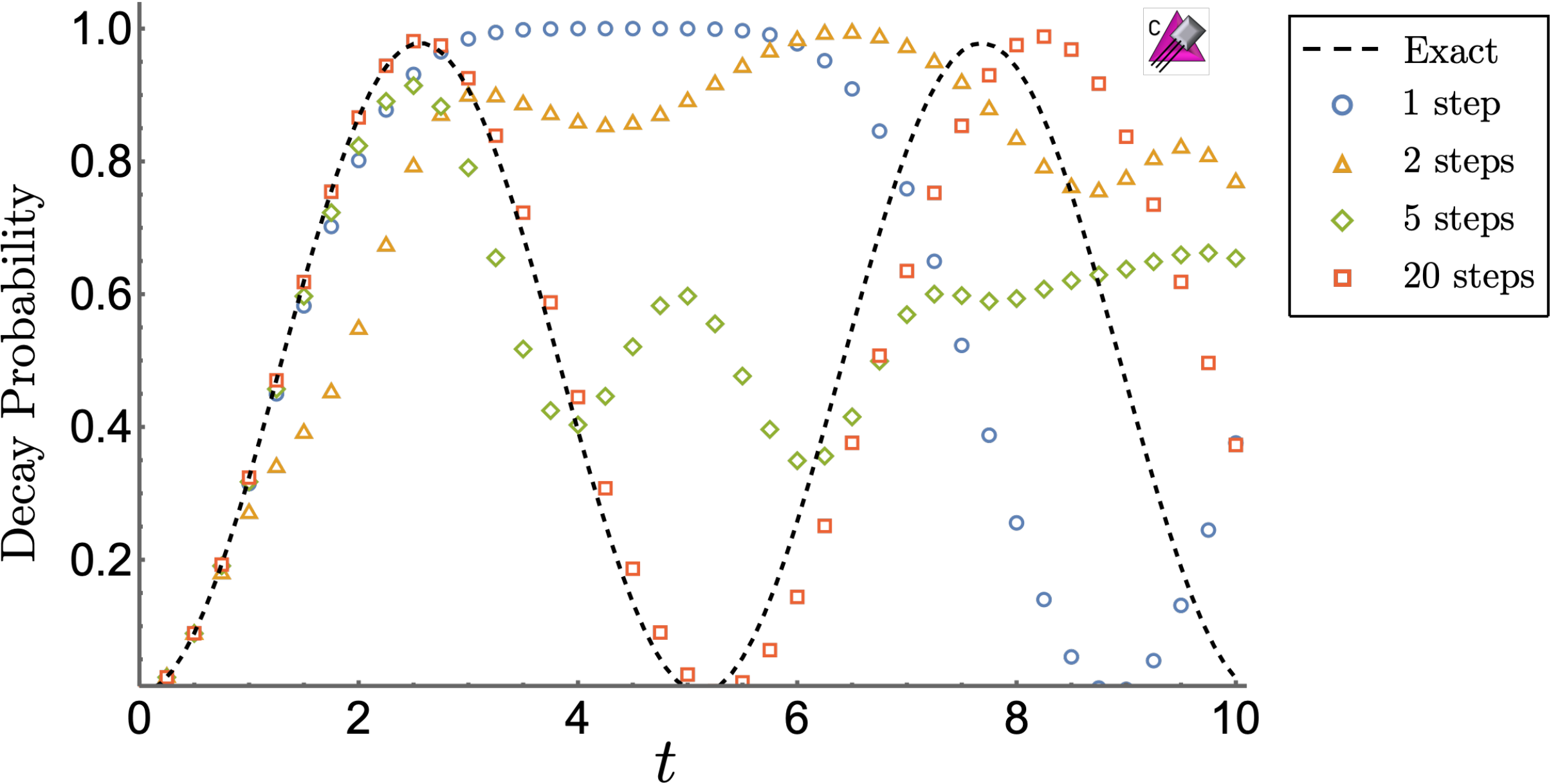}
    \caption{
The probability of $\beta$-decay, $\Delta^- \to \Delta^{0} + e + \overline{\nu}$, with $m_u = 0.9$, $m_d=2.1$, $m_{e,{\nu}} = 0$, $g=2$ and $G=0.5$ computed via exact diagonalization (dotted black line) and on the {\tt qiskit} quantum simulator~\cite{gadi_aleksandrowicz_2019_2562111} using $1,2,5,20$ Trotter steps.}
    \label{fig:BetaDecay}
\end{figure}

Once the $\Delta^-$ baryon state has been initialized 
on the register of qubits, it is then evolved in time with the full Hamiltonian.
The quantum circuits that implement the Trotterized time-evolution 
induced by $H_{{\rm quarks}}$ and $H_{{\rm glue}}$ were previously developed in
Ref.~\cite{Farrell:2022wyt}, 
where it was found that, by using an ancilla, each Trotter step 
can be implemented using 114 CNOTs. 
The lepton Hamiltonian, $\tilde{H}_{{\rm leptons}}$, has just single $Z$s which are Trotterized with single qubit rotations.
The circuits required to implement a Trotter step from $\tilde{H}_{\beta}$ are similar to those developed in Ref.~\cite{Farrell:2022wyt}, 
and their construction is outlined in App.~\ref{app:BetaCircuits}.
For the present choice of parameters, 
the main contribution to the initial ($\Delta^-$) wave function 
is $\lvert d_b d_g d_r \rangle$, 
i.e., the quark configuration associated with the ``bare" baryon in the d-sector and the trivial vacuum in the u-sector.
This implies that the dominant contribution to the $\beta$-decay 
is from the $\phi_0^{(u)\dagger} \phi_0^{(d)} \tilde{\chi}_0^{(e)\dagger} \tilde{\chi}_{1}^{(\nu)}$ 
term\footnote{Note that the $\phi_0^{(u)\dagger} \phi_0^{(d)} \tilde{\chi}_1^{(e)\dagger} \tilde{\chi}_{0}^{(\nu)}$ term is suppressed since the lepton vacuum in the tilde basis satisfies $\tilde{\chi}_1^{(e,\nu)\dagger} \lvert \Omega\rangle_{{\rm lep}} =  \tilde{\chi}_{0}^{(e,\nu)} \lvert \Omega\rangle_{{\rm lep}} = 0$.}  in Eq.~(\ref{eq:HbetaTilNoJW}), 
which acts only on valence quarks, and the $\beta$-decay operator can be approximated by 
\begin{equation}
    \tilde{H}_{\beta }^{{\rm val}}
    =
    \frac{G}{\sqrt{2}}
     \left (\sigma^-_{\overline{\nu}} \sigma^+_e  \sum_{c=r,g,b}\sigma^-_{d,c}Z^2 \sigma^+_{u,c}  + {\rm h.c.} \right  ) \ ,
     \label{eq:tildeBetaRed}
\end{equation}
for these parameter values. See App.~\ref{app:betaFull} for details on the validity of this approximation.
All of the results presented in this section implement this interaction, 
the Trotterization of which requires 50 CNOTs. 
Notice that, if the Trotterization of $\tilde{H}_{\beta}^{{\rm val}}$ is placed at the end of the first Trotter step, then\\
${U(t) = \exp(-i \tilde{H}_{\beta}^{{\rm val}} t) \times \exp \left [ -i (\tilde{H}_{{\rm leptons}} + H_{{\rm quarks}} + H_{{\rm glue}})t\right ]}$ and the initial exponential (corresponding to strong-interaction evolution) 
can be omitted as it acts on an eigenstate (the $\Delta^-$). 
This reduces the CNOTs required for one and two Trotter steps to $50$ and  $214$, respectively.
For an estimate of the number of CNOTs required to time evolve with the $\beta$-decay Hamiltonian on larger lattices see App.~\ref{app:LongJW}.
The probability of $\beta$-decay, 
as computed both through exact diagonalization of the Hamiltonian 
and through Trotterized time-evolution using the {\tt qiskit} classical simulator~\cite{gadi_aleksandrowicz_2019_2562111}, 
is shown in Fig.~\ref{fig:BetaDecay}.
The periodic structure is a finite volume effect, and the probability of 
$\beta$-decay is expected to tend to an exponential in time as $L$ increases, 
see App.~\ref{app:beta1p1aL}.

\begin{figure}[!ht]
    \centering
    \includegraphics[width=10cm]{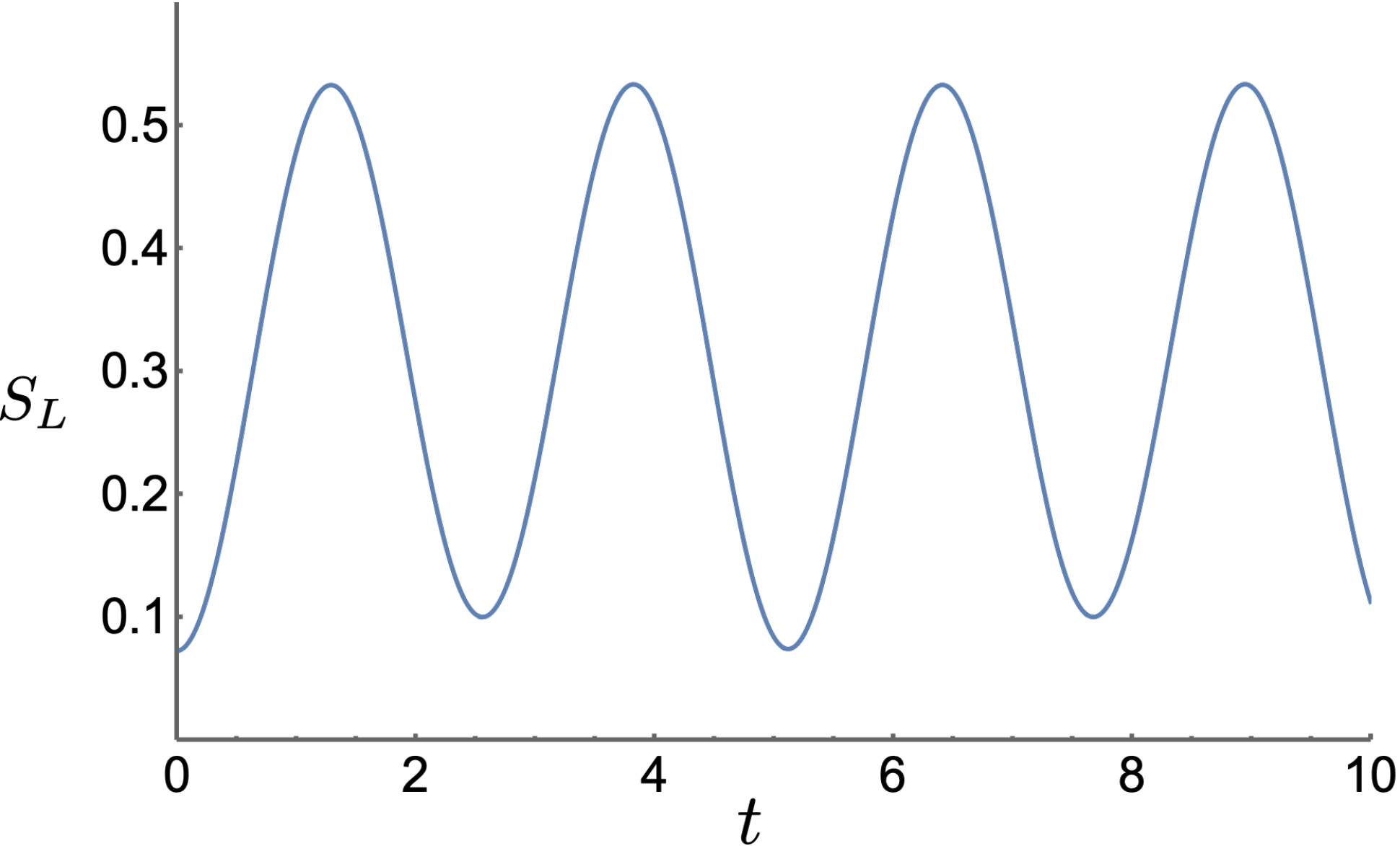}
    \caption{
    The linear entanglement entropy, $S_L$, between quarks and antiquarks plus leptons during the $\beta$-decay of an initial $\Delta^-$-baryon.}
    \label{fig:linEnt}
\end{figure}

Entanglement in quantum simulations of lattice gauge theories 
is a growing area of focus, 
see, e.g., Refs.~\cite{Ghosh:2015iwa,Soni:2015yga,Panizza:2022gvd,Rigobello:2021fxw}, and 
it is interesting to examine the evolution of entanglement during the $\beta$-decay process.
Before the decay, the quarks and antiquarks are together in a pure state as the leptons are in the vacuum, and subsequent time evolution of the state introduces components into the wavefunction that have non-zero population of the lepton states.
One measure of entanglement is the linear entropy,  
\begin{equation}
    S_L = 1 - \Tr[\rho_q^2]
    \ ,
\end{equation}
between the quarks and antiquarks plus leptons.
It is constructed by  tracing the full density matrix, $\rho$, 
over the antiquark and lepton sector to form the reduced density matrix 
$\rho_q = \Tr_{\overline{q}, {\rm leptons}} [\rho]$.
Figure~\ref{fig:linEnt} shows the linear entropy computed through exact diagonalization 
of the Hamiltonian with the parameters discussed previously in the text. 
By comparing with the persistence probability in Fig.~\ref{fig:BetaDecay}, 
it is seen that the entanglement entropy evolves at twice the frequency 
of the $\beta$-decay probability. 
This is because $\beta$-decay primarily transitions the baryon between the ground state of the $\Delta^-$ and $\Delta^0$. 
It is expected that these states will have a comparable amount of entanglement, 
and so the entanglement is approximately the same when the decay probabilities are $0$ and $1$. 
While this makes this particular example somewhat uninteresting, it does demonstrate that when multiple final states are accessible, the time-dependence of the entanglement structure might be revealing.

\subsection{Simulations Using Quantinuum's {\tt H1-1} 20 Qubit Trapped Ion Quantum Computer}
\label{sec:BetaSimB}
\noindent
Both the initial state preparation and one and two steps of Trotterized time evolution were executed
using Quantinuum's {\tt H1-1} 20 qubit trapped ion quantum computer~\cite{QuantHoney} and its simulator {\tt H1-1E}\footnote{The classical simulator {\tt H1-1E} includes depolarizing gate noise, leakage errors, crosstalk noise and dephasing noise due to transport and qubit idling~\cite{h1-1e}.} (for details on the specifications of {\tt H1-1}, see App.~\ref{app:H1specs}).
After transpilation onto the native gate set of {\tt H1-1}, a single Trotter step requires 59 $ZZ$ gates, while two Trotter steps requires 212 $ZZ$ gates.\footnote{The number of $ZZ$ gates could be further reduced by 5 by not resetting the ancilla.}
By post-selecting results on ``physical" states with baryon number $B=1$ and lepton number $L=0$ 
to mitigate single-qubit errors (e.g., Ref.~\cite{Klco:2019evd}),
approximately 90\% (50\%) of the total events from the one (two) Trotter step circuit remained. Additionally, for the two Trotter step circuit, results were selected where the ancilla qubit was in the $|0\rangle$ state (around 95\%).\footnote{For this type of error, the mid-circuit measurement and re-initialization option available for {\tt H1-1} could have been used to identify the case where the bit-flip occurred after the ancilla was used and the error had no effect on the final results.}

\begin{figure}[!t]
    \centering
    \includegraphics[width=\textwidth]{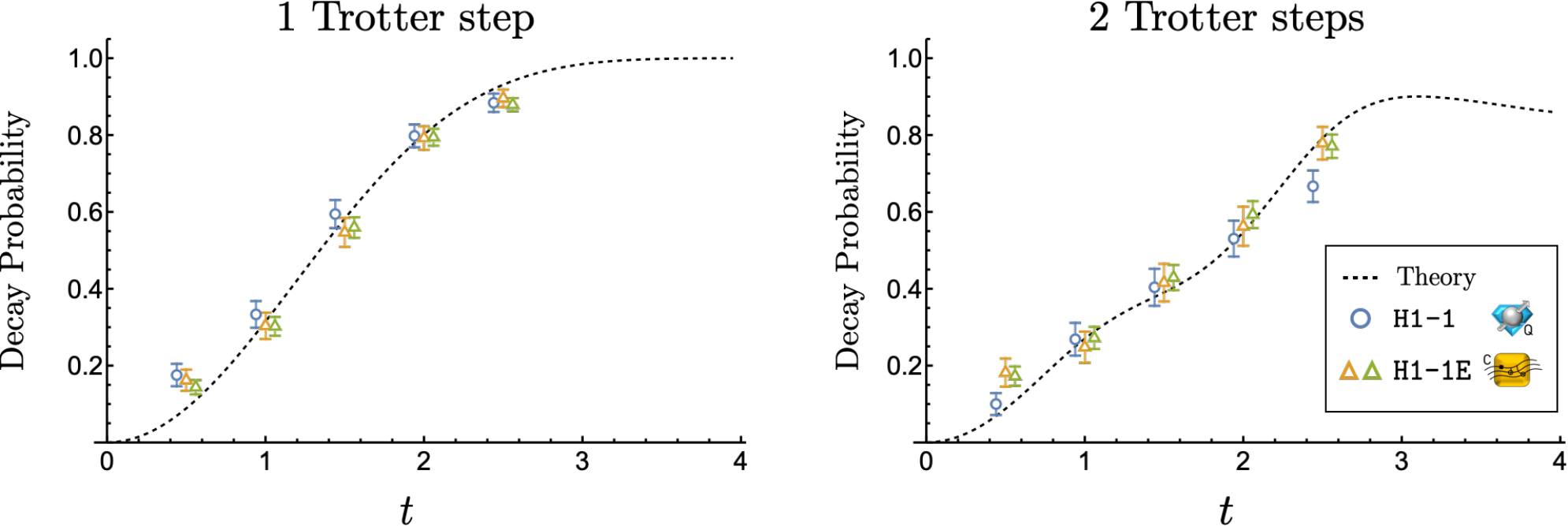}
    \caption{
    The probability of $\beta$-decay, $\Delta^- \to \Delta^{0} + e + \overline{\nu}$, with $m_u = 0.9$, $m_d=2.1$, $m_{e,{\nu}} = 0$, $g=2$ and $G=0.5$, using one (left panel) and two (right panel) Trotter steps (requiring 59 and 212 $ZZ$ gates, respectively), as given in Table~\ref{tab:H1results}.
    The dashed-black curves show the expected result from Trotterized time evolution, corresponding to the blue circles (orange triangles) in Fig.~\ref{fig:BetaDecay} for one (two) Trotter steps.
    The blue circles correspond to the data obtained on the {\tt H1-1} machine, 
    and the orange (green) triangles to the {\tt H1-1E} emulator, 
    each obtained from 200 shots (400 shots). The points have been shifted slightly along the $t$-axis for clarity.
    Error mitigation beyond physical-state post-selection has not been performed.
    The weak Hamiltonian in the time-evolution responsible for the decay is given in Eq.~(\ref{eq:tildeBetaRed}).
    }
    \label{fig:BetaDecayH1}
\end{figure}

The results of the simulations 
are shown in Fig.~\ref{fig:BetaDecayH1} and given in Table~\ref{tab:H1results}. 
By comparing the results from {\tt H1-1} and {\tt H1-1E} (using 200 shots) it is seen that the simulator is able to faithfully reproduce the behavior of the quantum computer.
The emulator was also run with 400 shots and clearly shows convergence to the expected value, verifying that the agreement between data and theory was not an artifact due to low statistics (and large error bars). 
Compared with the results presented in Ref.~\cite{Farrell:2022wyt} that were performed using IBM's {\tt ibmq$\_$jakarta} and {\tt ibm$\_$perth}, 
error mitigation techniques were not 
applied to the present simulations due to the overhead in resource requirements.
Specifically, Pauli twirling, dynamical decoupling, decoherence renormalization and measurement error mitigation 
were not performed. This is practical because the two-qubit gate, state preparation and measurement (SPAM) errors are an order of magnitude smaller on Quantinuum's trapped ion system 
compared to those of IBM's superconducting qubit systems (and a similar error rate on the single-qubit gates)~\cite{Pelofske:2022vyy}. 
\begin{table}[!ht]
\renewcommand{\arraystretch}{1.2}
\begin{tabularx}{0.85\textwidth}{||c | Y | Y | Y | Y | Y | Y | Y | Y ||}
\hline
\multicolumn{9}{||c||}{
Single-Baryon Decay Probabilities using Quantinuum's {\tt H1-1} and {\tt H1-1E}} \\
\hline
 & \multicolumn{4}{c|}{1 Trotter step} & \multicolumn{4}{c||}{2 Trotter steps} \\
 \hline
 $t$ & 
 {\tt H1-1} & {\tt H1-1E} & \makecell{{\tt H1-1E} \\($\times 2$ stats)} & Theory & 
 {\tt H1-1} & {\tt H1-1E} & \makecell{{\tt H1-1E} \\($\times 2$ stats)} & Theory
 \\
 \hline
 0.5 & 
 0.175(29) & 0.162(28) & 0.144(19) & 0.089 & 
 0.100(29) & 0.182(37) & 0.173(25) & 0.088\\
 \hline
 1.0 & 
 0.333(35) & 0.303(34) & 0.302(25) & 0.315 & 
 0.269(43) & 0.248(41) & 0.272(29) & 0.270 \\
 \hline
 1.5 & 
 0.594(37) & 0.547(38) & 0.559(27) & 0.582 & 
 0.404(48) & 0.416(49) & 0.429(33)  & 0.391 \\
 \hline
 2.0 & 
 0.798(30) & 0.792(30) & 0.794(22) & 0.801 & 
 0.530(47) & 0.563(51) & 0.593(35) & 0.547 \\
 \hline
 2.5 & 
 0.884(24) & 0.896(23) & 0.879(17) & 0.931 & 
 0.667(41) & 0.779(43) & 0.771(30) & 0.792 \\
 \hline
\end{tabularx}
\renewcommand{\arraystretch}{1}
\caption{
The probability of $\beta$-decay, 
$\Delta^- \to \Delta^{0} + e + \overline{\nu}$, on $L=1$ spatial lattice with $m_u = 0.9$, $m_d=2.1$, $m_{e,{\nu}} = 0$, $g=2$ and $G=0.5$.
These simulations
were performed using Quantinuum's {\tt H1-1} and {\tt H1-1E} and included the initial state preparation and subsequent time evolution under 1 and 2 Trotter steps. The results are displayed in Fig.~\ref{fig:BetaDecayH1}. 
The columns labeled ($\times 2$ stats) were obtained using 400 shots, compared to the rest, that used 200 shots, and
uncertainties were computed assuming the results follow a binomial distribution.
}
\label{tab:H1results}
\end{table}
%

\section{Speculation about Quantum Simulations with a Hierarchy of Length Scales}
\noindent
It is interesting to consider how a hierarchy of length scales, 
as present in the SM, may be helpful in error correction.
In the system we have examined, the low energy strong sector is composed of mesons, baryons and nuclei, with both color singlet and non-singlet excitations (existing at higher energies).
As observed in Ref.~\cite{Farrell:2022wyt}, OBCs allow for 
relatively low-energy colored ``edge" states to exist near the boundary of the lattice.
The energy of a color non-singlet grows linearly with its distance 
from the boundary, leading to a force on colored objects.
This will cause colored errors in the bulk to migrate to the edge of the lattice where they could be detected and possibly removed.
This is one benefit of using axial gauge, where Gauss's law is automatically enforced, 
and a colored ``error" in the bulk generates a color flux tube that extends to the boundary.

Localized two-bit-flip errors  can create color-singlet 
excitations that do not experience a force toward the boundary, but which are
vulnerable to weak decay. 
For sufficiently large lattices,  color singlet excitations will decay weakly down to stable states
enabled by the near continuum of lepton states. 
In many ways, this resembles the
quantum imaginary-time evolution (QITE)~\cite{Kamakari:2021nmf,Hubisz:2020vhx,Turro:2021vbk}
algorithm, which is a special case of coupling to open systems,
where quantum systems are driven into their ground state by embedding them in a larger system that acts as a heat reservoir.
One can speculate that, in the future, quantum simulations of QCD 
will benefit from also including electroweak interactions as a mechanism to cool the strongly-interacting sector from particular classes of errors.

This particular line of investigation is currently at a ``schematic'' level, and significantly more work is required to quantify its utility.
Given the quantum resource requirements, it is likely that the Schwinger model will 
provide a suitable system to explore such scenarios.

\section{Summary and Conclusions}
\noindent
Quantum simulations of SM physics is in its infancy and, for practical reasons, has been previously
limited to either QCD or QED in one or two spatial dimensions. 
In this work, we have started the integration of the electroweak sector into quantum simulations of QCD by examining the time-evolution of the $\beta$-decay of one baryon.
In addition to the general framework that allows for 
simulations of arbitrary numbers of lattice sites in one dimension, 
we present results for $L=1$ spatial lattice site, which requires 16 qubits.
Explicitly, this work considered quantum simulations of 
$\Delta^-\rightarrow\Delta^0 e \overline{\nu}$
in two flavor $1+1$D QCD for $L=1$ spatial lattice site.
Simulations were performed using Quantinuum's {\tt H1-1} 20-qubit trapped ion quantum computer
and classical simulator {\tt H1-1E}, 
requiring 17 (16 system and 1 ancilla) qubits. 
Results were presented for both one and two Trotter steps, including the state preparation of the initial baryon, requiring 59 and 212 two-qubit gates respectively.
Even with 212 two-qubit gates, {\tt H1-1} provided results that 
are consistent with the expected results, within uncertainties, without error-mitigation beyond physical-state post selection.  
While not representative of $\beta$-decay in the continuum, 
these results demonstrate the potential of quantum simulations to determine 
the real-time evolution of decay and reaction processes in nuclear and 
high-energy processes.
High temporal-resolution studies of the evolution of the quarks and gluons
during hadronic decays and nuclear reactions
are expected to provide new insights into the mechanisms responsible for these processes, 
and lead to new strategies for further reducing systematic errors in their prediction.

\begin{acknowledgements}
\noindent
We would like to thank 
Silas Beane, Natalie Klco and Randy Lewis for helpful discussions and comments.
This work was supported, in part, by 
the U.S. Department of Energy grant DE-FG02-97ER-41014 (Farrell), 
the U.S.\ Department of Energy,
Office of Science, Office of Nuclear Physics, InQubator for Quantum Simulation (IQuS) 
(\url{https://iqus.uw.edu})
under Award Number DOE
(NP) Award DE-SC0020970 (Chernyshev, Farrell, Powell, Savage, Zemlevskiy), 
and the 
Quantum Science Center (QSC)
(\url{https://qscience.org}), 
a National Quantum Information Science Research Center of the U.S.\ Department of Energy (DOE) (Illa).
This work is also supported, in part, through the Department of Physics 
(\url{https://phys.washington.edu}) 
and the College of Arts and Sciences 
(\url{https://www.artsci.washington.edu})
at the University of Washington.
This research used resources of the Oak Ridge Leadership Computing Facility, which is a DOE Office of Science User Facility supported under Contract DE-AC05-00OR22725.
We have made extensive use of Wolfram {\tt Mathematica}~\cite{Mathematica},
{\tt python}~\cite{python3,Hunter:2007}, {\tt julia}~\cite{Julia-2017},
{\tt jupyter} notebooks~\cite{PER-GRA:2007} 
in the {\tt Conda} environment~\cite{anaconda},
and the quantum programming environments: Google's {\tt cirq}~\cite{cirq_developers_2022_6599601}, IBM's {\tt qiskit}~\cite{gadi_aleksandrowicz_2019_2562111}, 
and CQC's {\tt pytket}~\cite{sivarajah2020t}.
\end{acknowledgements}

\clearpage
\appendix


\section{The Complete Spin Hamiltonian for \texorpdfstring{$L=1$}{L=1}}
\label{app:fullHam}
\noindent
After the JW mapping of the Hamiltonian to qubits, and using the tilde-basis for the leptons, 
the four contributing terms are
\begin{subequations}
    \label{eq:H2flavL1}
    \begin{align}
    H = & \ H_{{\rm quarks}}\ +\ \tilde{H}_{{\rm leptons}}\ +\ H_{{\rm glue}} \ +\ \tilde{H}_{\beta} ,\\[4pt]
    H_{{\rm quarks}}= & \ \frac{1}{2} \left [ m_u\left (Z_0 + Z_1 + Z_2 -Z_6 - Z_7 - Z_8 + 6\right )+ m_d\left (Z_3 + Z_4 + Z_5 -Z_9 - Z_{10} - Z_{11} + 6\right ) \right ] \nonumber \\
    &-\, \frac{1}{2} (\sigma^+_6 Z_5 Z_4 Z_3 Z_2 Z_1 \sigma^-_0 + \sigma^-_6 Z_5 Z_4 Z_3 Z_2 Z_1 \sigma^+_0 + \sigma^+_7 Z_6 Z_5 Z_4 Z_3 Z_2 \sigma^-_1 + \sigma^-_7 Z_6 Z_5 Z_4 Z_3 Z_2 \sigma^+_1  \nonumber \\
    &+\, \sigma^+_8 Z_7 Z_6 Z_5 Z_4 Z_3 \sigma^-_2 + \sigma^-_8 Z_7 Z_6 Z_5 Z_4 Z_3 \sigma^+_2 + \sigma^+_9 Z_8 Z_7 Z_6 Z_5 Z_4 \sigma^-_3 + \sigma^-_9 Z_8 Z_7 Z_6 Z_5 Z_4 \sigma^+_3 \nonumber \\
    &+\, \sigma^+_{10} Z_9 Z_8 Z_7 Z_6 Z_5 \sigma^-_4 + \sigma^-_{10} Z_9 Z_8 Z_7 Z_6 Z_5 \sigma^+_4 + \sigma^+_{11} Z_{10} Z_9 Z_8 Z_7 Z_6 \sigma^-_5 + \sigma^-_{11} Z_{10} Z_9 Z_8 Z_7 Z_6 \sigma^+_5 ) \ ,
        \label{eq:Hkin2flavL1}\\[4pt]
    \tilde{H}_{{\rm leptons}} = & \ \frac{1}{4}\sqrt{1 +4 m_e^2}(Z_{13} - Z_{15}) + \frac{1}{4}\sqrt{1 +4 m_{\nu}^2}(Z_{12} - Z_{14}) \nonumber \\[4pt]
    H_{{\rm glue}} = & \ \frac{g^2}{2} \bigg [ \frac{1}{3}(3 - Z_1 Z_0 - Z_2 Z_0 - Z_2 Z_1) + \sigma^+_4\sigma^-_3\sigma^-_1\sigma^+_0  + \sigma^-_4\sigma^+_3\sigma^+_1\sigma^-_0  + \sigma^+_5Z_4\sigma^-_3\sigma^-_2Z_1\sigma^+_0 + \sigma^-_5Z_4\sigma^+_3\sigma^+_2Z_1\sigma^-_0 \nonumber \\
    & +\, \sigma^+_5\sigma^-_4\sigma^-_2\sigma^+_1 + \sigma^-_5\sigma^+_4\sigma^+_2\sigma^-_1  \nonumber \\
    & +\,\frac{1}{12}\left (2 Z_3 Z_0 + 2Z_4 Z_1 + 2Z_5 Z_2 - Z_5 Z_0 - Z_5 Z_1 - Z_4 Z_2 - Z_4 Z_0 - Z_3 Z_1  - Z_3 Z_2 \right ) \bigg ] \ ,
    \label{eq:Hel2flavL1}\\[4pt] 
    \tilde{H}_{\beta} = & \ \frac{G}{\sqrt{2}}  \bigg \{ \frac{1}{2}( s_+^e s_+^{\nu} + s_-^e s_-^{\nu})\big [(\textcolor{blue}{\sigma^-_{14} \sigma^+_{13}} -  \sigma^+_{15} Z_{14} Z_{13}\sigma^-_{12})\big (\textcolor{blue}{\sigma^-_{3} Z_2 Z_1 \sigma^+_0 + \sigma^-_{4} Z_3 Z_2 \sigma^+_1 + \sigma^-_5 Z_4 Z_3 \sigma^+_2} + \sigma^-_{9} Z_8 Z_7 \sigma^+_6  \nonumber \\
    &+ \ \sigma^-_{10} Z_9 Z_8 \sigma^+_7 + \sigma^-_{11} Z_{10} Z_9 \sigma^+_8) \ + \ (\textcolor{blue}{\sigma^+_{14} \sigma^-_{13}} - \sigma^-_{15} Z_{14} Z_{13}\sigma^+_{12})\big (\textcolor{blue}{\sigma^+_{3} Z_2 Z_1 \sigma^-_0 + \sigma^+_{4} Z_3 Z_2 \sigma^-_1 + \sigma^+_5 Z_4 Z_3 \sigma^-_2}  \nonumber \\
    & + \ \sigma^+_{9} Z_8 Z_7 \sigma^-_6 + \sigma^+_{10} Z_9 Z_8 \sigma^-_7  + \sigma^+_{11} Z_{10} Z_9 \sigma^-_8)\big ]  \nonumber \\ 
    &- \ \frac{1}{2}(s_+^e s_-^{\nu}  +  s_-^e s_+^{\nu}) \big [ (\sigma^-_{14} \sigma^+_{13} +  \sigma^+_{15} Z_{14} Z_{13}\sigma^-_{12}) \big( \sigma^-_{9} Z_8 Z_7 Z_6 Z_5 Z_4 Z_3 Z_2 Z_1 \sigma^+_0 + \sigma^-_{10} Z_9 Z_8 Z_7 Z_6 Z_5 Z_4 Z_3 Z_2 \sigma^+_1 \nonumber \\
    &+ \ \sigma^-_{11} Z_{10} Z_9 Z_8 Z_7 Z_6 Z_5 Z_4 Z_3 \sigma^+_2 + \sigma^+_{6} Z_5 Z_4 \sigma^-_3 + \sigma^+_{7} Z_6 Z_5 \sigma^-_4 + \sigma^+_{8} Z_7 Z_6 \sigma^-_5 \big )\nonumber \\
    &+ \ (\sigma^+_{14} \sigma^-_{13} + \sigma^-_{15} Z_{14} Z_{13}\sigma^+_{12}) \big( \sigma^+_{9} Z_8 Z_7 Z_6 Z_5 Z_4 Z_3 Z_2 Z_1 \sigma^-_0 + \sigma^+_{10} Z_9 Z_8 Z_7 Z_6 Z_5 Z_4 Z_3 Z_2 \sigma^-_1 \nonumber \\
    &+ \  \sigma^+_{11} Z_{10} Z_9 Z_8 Z_7 Z_6 Z_5 Z_4 Z_3 \sigma^-_2 + \sigma^-_{6} Z_5 Z_4 \sigma^+_3 + \sigma^-_{7} Z_6 Z_5 \sigma^+_4 + \sigma^-_{8} Z_7 Z_6 \sigma^+_5 \big ) \big ] \bigg \} \ .
    \end{align}
\end{subequations}
In the mapping, the qubits are indexed right-to-left and, 
for example, qubit zero (one) corresponds to a red (green) up-quark.
The terms highlighted in blue provide the leading contribution to the $\beta$-decay process 
for the parameters used in the text and make up the operator used for the simulations performed in Sec.~\ref{sec:BetaSim}.

\section{\texorpdfstring{$\beta$}{Beta}-Decay in the Standard Model}
\label{app:betaSM}
\noindent 
To put our simulations in $1+1$D into context, 
it is helpful to  outline relevant aspects of single-hadron $\beta$-decays 
in the SM in $3+1$D.
Far below the electroweak symmetry-breaking scale,  
charged-current interactions can be included as an infinite 
set of effective operators in a systematic EFT description, ordered by their contributions in powers of low-energy scales divided by appropriate powers of $M_W$.
For instance, $\beta$-decay rates between hadrons scale as 
$\sim \Lambda (G_F \Lambda^2 )^2 (\Lambda / M_W )^n$, 
where $\Lambda$ denotes the low-energy scales,
$\frac{G_F}{\sqrt{2}} = \frac{g_2^2}{8 M_W^2}$ is Fermi's constant and 
LO (in $\Lambda / M_W$)
corresponds to $n=0$.
By matching operators at LO in SM interactions, the $\beta$-decay of the neutron is induced by an effective Hamiltonian density of the form~\cite{Feynman:1958ty,Sudarshan:1958vf}
\begin{equation}
    {\cal H}_\beta = 
    \frac{G_F}{\sqrt{2}} \ V_{ud} \ 
    \overline{\psi}_u\gamma^\mu (1-\gamma_5)\psi_d\ 
    \overline{\psi}_e\gamma_\mu (1-\gamma_5)\psi_{\nu_e}  
    \ +\ {\rm h.c.}
    \ ,
    \label{eq:betaHamiSM}
\end{equation}
where $V_{ud}$ is the element of the CKM matrix for $d\rightarrow u$ transitions.
As ${\cal H}_\beta $ factors into contributions from lepton and quark operators, the matrix element factorizes into a plane-wave lepton contribution and a non-perturbative hadronic component requiring matrix elements of the quark operator between hadronic states. 
With the mass hierarchies and symmetries in nature, 
there are two dominant form factors, so that,
\begin{equation}
    \langle p(p_p) | \overline{\psi}_u\gamma^\mu (1-\gamma_5)\psi_d | n(p_n)\rangle = 
    \overline{U}_p \left[\ g_V(q^2) \gamma^\mu - g_A(q^2) \gamma^\mu \gamma_5\ \right] U_n \ ,
\end{equation}
where $q$ is the four-momentum transfer of the process, $g_V(0) = 1$ in the isospin limit and $g_A(0)=1.2754(13)$~\cite{Workman:2022ynf} as measured in experiment.
The matrix element for $n\rightarrow p e^- \overline{\nu}_e$
calculated from the Hamiltonian in Eq.~(\ref{eq:betaHamiSM})
is
\begin{equation}
    \lvert \mathcal{M} \rvert^2 = 16 G_F^2 \lvert V_{ud}\rvert^2 M_n M_p (g_V^2 + 3 g_A^2)(E_{\nu}E_e + \frac{g_V^2-g_A^2}{g_V^2 + 3 g_A^2} {\bf p}_e \cdot {\bf p}_{\nub}) \ ,
\end{equation}
which leads to a neutron width of 
(at LO in $(M_n-M_p)/M_n$ and $m_e/M_n$)
\begin{equation}
    \Gamma_n =
    \frac{G_F^2 |V_{ud}|^2 (M_n-M_p)^5}{60\pi^3}
    \ \left( g_V^2 + 3 g_A^2 \right)\ f^\prime(y) \ ,
\end{equation}
where $f^\prime(y)$ is a phase-space factor, 
\begin{equation}
    f^\prime (y) = \sqrt{1-y^2}\left(1 - \frac{9}{2}y^2 - 4 y^4\right)
    \ -\ \frac{15}{2}y^4 \log\left[ \frac{y}{\sqrt{1-y^2}+1}\right] \ ,
\end{equation}
and $y=m_e/(M_n-M_p)$.
Radiative effects, recoil effects and other higher-order contributions have been neglected.

\section{\texorpdfstring{$\beta$}{Beta}-Decay in \texorpdfstring{$1+1$}{1+1} Dimensions: The \texorpdfstring{$L=\infty$}{L to infinity} and Continuum Limits}
\label{app:beta1p1}
\noindent 
In $1+1$D, the fermion field has dimensions 
$\left[\psi\right] = \frac{1}{2}$, 
and a four-Fermi operator has dimension 
$[\hat \theta ] = 2$.  
Therefore, while in $3+1$D 
$\left[G_F \right] = -2$, 
in $1+1$D, the coupling has dimension $\left[G \right] = 0$.
For our purposes, to describe the $\beta-$decay of a $\Delta^-$-baryon in $1+1$D,
we have chosen to work with an effective Hamiltonian of the form
\begin{equation}
    {\cal H}_\beta^{1+1} \ = \ 
    \frac{G}{\sqrt{2}} \
    \overline{\psi}_u\gamma^\mu \psi_d\ 
    \overline{\psi}_e\gamma_\mu\psi_{\overline{\nu}} 
    \ +\ {\rm h.c.}
    \ =\ 
    \frac{G}{\sqrt{2}} \
    \overline{\psi}_u\gamma^\mu \psi_d\ 
    \overline{\psi}_e\gamma_\mu \mathcal{C} \psi_{\nu} 
        \ +\ {\rm h.c.}
\ ,
\label{eq:betaHami1p1}
\end{equation}
where we have chosen the basis
\begin{equation}
    \gamma_0 \ = \ 
    \left(
    \begin{array}{cc}
    1&0 \\ 0&-1
    \end{array}
    \right)
    \ \ ,\ \ 
    \gamma_1 \ = \ 
    \left(
    \begin{array}{cc}
    0&1 \\ -1&0
    \end{array}
    \right)
    \ =\ \mathcal{C}
    \ ,\ 
    \gamma_0\gamma_\mu^\dagger \gamma_0\ =\ \gamma_\mu
    \ ,\ 
    \gamma_0 \mathcal{C} ^\dagger \gamma_0\ =\ \mathcal{C}
    \ \ ,\ \ 
    \{\gamma_\mu , \gamma_\nu \} \ =\ 2 g_{\mu\nu}
    \ .
\label{eq:gammaMats1p1}
\end{equation}
For simplicity, the CKM matrix element is set equal to unity 
as only one generation of particles is considered.

In the limit of exact isospin symmetry, which we assume to be approximately valid in this appendix, 
the four $\Delta$ baryons form an isospin quartet 
and can be embedded in a tensor $T^{abc}$ (as is the case for the $\Delta$ resonances in nature)
as 
$T^{111}=\Delta^{++}$,
$T^{112}=T^{121}=T^{211}=\Delta^{+}/\sqrt{3}$,
$T^{122}=T^{221}=T^{212}=\Delta^{0}/\sqrt{3}$,
$T^{222}=\Delta^{-}$.
Matrix elements of the isospin generators
are reproduced by an effective operator of the form
\begin{equation}
    \overline\psi_q \gamma^\mu \tau^\alpha \psi_q \ \rightarrow \
    3 \overline{T}_{abc} \gamma^\mu \left(\tau^\alpha \right)^c_d T^{abd} \ ,
\label{eq:DeltaIgens}
\end{equation}
which provides a Clebsch-Gordan coefficient for isospin raising operators,
\begin{equation}
    \overline\psi_q \gamma^\mu \tau^+ \psi_q \ \rightarrow \
    \sqrt{3}\  \overline{\Delta^{++}}\gamma^\mu\Delta^+ 
    \ +\ 2 \ \overline{\Delta^{+}}\gamma^\mu\Delta^0
    \ +\ 
    \sqrt{3}\  \overline{\Delta^{0}}\gamma^\mu\Delta^- \ .
\label{eq:DeltaIgens2}
\end{equation}

The matrix element for $\beta$-decay factorizes at LO in the electroweak interactions.
The hadronic component of the matrix element is given by 
\begin{align}
\langle \Delta^0(p_0) | \overline{\psi}_u\gamma^\alpha \psi_d | \Delta^-(p_-)\rangle & = \sqrt{3} g_V(q^2) \ \overline{U}_{\Delta^0}  \gamma^\alpha  U_{\Delta^-} \ =\ H^\alpha \ , \nonumber\\
H^\alpha H^{\beta\ \dagger} & = 3 |g_V(q^2) |^2 {\rm Tr}\left[ \gamma^\alpha \left( \pslash_- + M_{\Delta^-}\right) \gamma^\beta 
\left( \pslash_0 + M_{\Delta^0}\right) \right] \nonumber\\
& = 6 |g_V(q^2) |^2  
\left[p_-^\alpha p_0^\beta  + p_0^\alpha p_-^\beta  - g^{\alpha\beta} (p_-\cdot p_0)
\ +\ M_{\Delta^-} M_{\Delta^0} g^{\alpha\beta}\right] \ =\ H^{\alpha\beta} \ ,
\end{align}
and the leptonic component of the matrix element is given by, assuming that the electron and neutrino are massless, 
\begin{align}
\langle e^- \overline{\nu}_e | \overline{\psi}_e \gamma^\alpha C \psi_{\nu}  | 0\rangle 
& = \overline{U}_e\gamma^\alpha C V_\nu\ =\ L^\alpha \ ,
\nonumber\\
L^\alpha L^{\beta\ \dagger} & = {\rm Tr}\left[\ 
\gamma^\alpha C \pslash_\nu C \gamma^\beta \pslash_e \right] \ =\  {\rm Tr}\left[\ \gamma^\alpha \overline{\pslash}_\nu \gamma^\beta \pslash_e \right] \nonumber\\
& = 
2 \left[ \overline{p}_\nu^\alpha p_e^\beta + \overline{p}_\nu^\beta p_e^\alpha  -  g^{\alpha\beta} (\overline{p}_\nu\cdot p_e) \right] \ =\ L^{\alpha\beta}\ ,
\end{align}
where $p = (p^0, +p^1)$ and 
$\overline{p}=(p^0, -p^1)$.
Therefore, the squared matrix element of the process is
\begin{equation}
    |{\cal M}|^2 \ = \
    \frac{G^2}{2}
    H^{\alpha\beta} L_{\alpha\beta} \ = \ 
    12 G^2 g_V^2  M_{\Delta^-} 
    \left( M_{\Delta^-} - 2 E_{\overline{\nu}}\right)
    \left( E_e E_{\overline{\nu}} - {\bf p}_e\cdot {\bf p}_{\overline{\nu}} \right) \ ,
\end{equation}
from which
the delta decay width can be determined by standard methods,
\begin{align}
\Gamma_{\Delta^-} & =
\frac{1}{2M_{\Delta^-}}
\int 
\frac{d{\bf p}_e}{4\pi E_e}
\frac{d{\bf p}_{\overline{\nu}}}{4\pi E_{\overline{\nu}}}
\frac{d{\bf p}_0}{4\pi E_0} (2\pi)^2 \delta^2(p_- - p_0 - p_e - p_{\overline{\nu}})
|{\cal M}|^2 
\nonumber\\
& =
3 \frac{G^2 g_V^2}{2\pi}
\int\  dE_e\  dE_{\overline{\nu}}\ \delta(Q - E_e - E_{\overline{\nu}})
\ +\ {\cal O}\left(Q^n/M_\Delta^n\right)
\nonumber\\
& =
3 \frac{G^2 g_V^2 Q}{2\pi}
\ +\ {\cal O}\left(Q^n/M_\Delta^n\right)
\ ,
\label{eq:1p1decaywidth}
\end{align}
where $Q= M_{\Delta^-} - M_{\Delta^0}$ and we have retained only the leading terms in
an expansion in $Q/M_\Delta$ and evaluated the vector form factor at $g_V(q^2=0) \equiv g_V$. 
The electron and neutrino masses have been set to zero, and the inclusion of non-zero masses will lead to a phase-space factor, $f_1$, 
reducing the width shown in Eq.~(\ref{eq:1p1decaywidth}), 
and which becomes $f_1=1$ in the massless limit.

\section{\texorpdfstring{$\beta$}{Beta}-Decay in \texorpdfstring{$1+1$}{1+1} Dimensions: Finite \texorpdfstring{$L$}{L} and Non-zero Spatial Lattice Spacing}
\label{app:beta1p1aL}
\noindent 
The previous appendix computed the $\beta$-decay rate 
in $1+1$D in infinite volume and in the continuum.
However, lattice  calculations of such processes will necessarily be performed with a non-zero lattice spacing and a finite number of lattice points.
For calculations done on a Euclidean-space lattice, significant work has been done to develop the machinery used to extract physically meaningful results. 
This formalism was initially pioneered by L\"{u}scher~\cite{Luscher:1985dn,Luscher:1986pf,Luscher:1990ux} for hadron masses and two-particle scattering, and has been extended to more complex systems relevant to electroweak processes (Lellouch-L\"{u}scher)~\cite{Lellouch:2000pv,Detmold:2004qn,Christ:2005gi,Kim:2005gf,Hansen:2012tf,Meyer:2011um,Briceno:2012yi,Feng:2014gba,Briceno:2012yi,Meyer:2012wk,Bernard:2012bi,Agadjanov:2014kha,Briceno:2014uqa,Briceno:2015csa,Briceno:2015tza,Briceno:2019opb,Briceno:2020vgp} and 
to nuclear physics~\cite{Beane:2003da,Detmold:2004qn,Beane:2007qr,Beane:2007es,Luu:2010hw,Luu:2011ep,Davoudi:2011md,Meyer:2012wk,Briceno:2012rv,Briceno:2013lba,Briceno:2013bda,Briceno:2013hya,Briceno:2014oea,Grabowska:2021qqz}.   
L\"{u}scher's method was originally derived from an analysis of Hamiltonian dynamics in Euclidean space and later from a field theoretic point of view directly from correlation functions. 
The challenge is working around the Maiani-Testa theorem~\cite{Maiani:1990ca} and reliably determining Minkowski-space matrix elements from Euclidean-space observables.
This formalism has been used successfully for a number of important quantities, and continues to be the workhorse for Euclidean-space computations.

As quantum simulations provide observables directly in Minkowski space,
understanding the finite-volume and non-zero lattice spacing artifacts requires a similar but different analysis than in Euclidean space.\footnote{Estimates of such effects in model 1+1 dimensional simulations can be found in Ref.~\cite{PhysRevD.103.014506}.}  
While the method used in Euclidean space of determining S-matrix elements for scattering processes from energy eigenvalues can still be applied, Minkowski space simulations will also allow for a direct evaluation of scattering processes, removing some of the modeling that remains in Euclidean-space calculations.\footnote{
For example, the energies of states in different volumes are different, 
and so the elements of the scattering matrix are constrained over 
a range of energies and not at one single energy,
and {\it a priori} unknown interpolations are modeled.
}
Neglecting electroweak interactions beyond $\beta$-decay means that the final state leptons are non-interacting (plane-waves when using periodic boundary conditions),
and therefore the modifications to the density of states due to interactions, as encapsulated within the L\"{u}scher formalism, are absent.

With Hamiltonian evolution of a system described within a finite-dimensional Hilbert space, the persistence amplitude of the initial state coupled to final states via the weak Hamiltonian will be determined by the sum over oscillatory amplitudes.  
For a small number of final states, the amplitude will return to unity after some finite period of time.  
As the density of final states near the energy of the initial state becomes large, there will be cancellations among the oscillatory amplitudes, and the persistence probability will begin to approximate the ``classic" exponential decay over some time interval.  
This time interval will extend  to infinity as the density of states tends to a continuous spectrum. 
It is important to understand how to reliably extract an estimate of the decay rate, with a quantification of systematic errors, from the amplitudes measured in a quantum simulation.  
This is the subject of future work, but here a simple model will be used to demonstrate some of the relevant issues.

\begin{figure}[!ht]
    \centering
    \includegraphics[width=12cm]{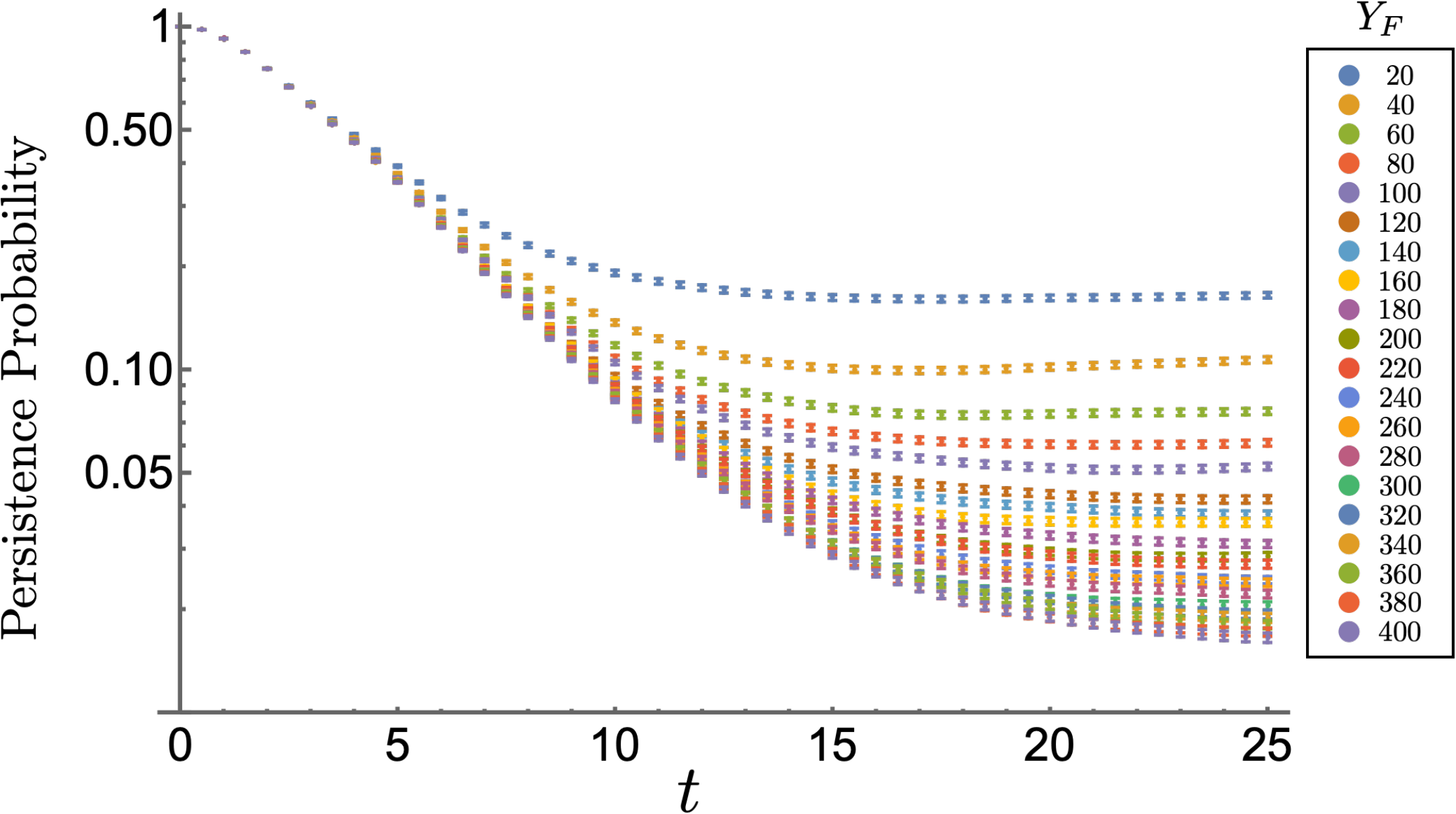}
    \caption{
    Ensemble averages (over 2000 random samples) of the persistence probability of an initial state in one sector of a strong Hamiltonian undergoing weak decay to states in a different sector, as described in this appendix. 
    The different colored points are results from calculations with an increasing number of final states, $Y_F$. 
    The weak coupling scales so that the decay probability converges to a well-defined value as the density of final states tends to a continuum.
    }
    \label{fig:ExpDecay}
\end{figure}

Consider the weak decay of a strong eigenstate in one sector to a strong eigenstate in a different sector (a sector is defined by its strong quantum numbers).
For this demonstration, 
we calculate the persistence probability of the initial state, averaged over random weak and strong Hamiltonians and initial states, as the number of states  below a given energy increases (i.e. increasing density of states).
Concretely, the energy eigenvalues of the initial strong sector range from 0 to 1.1, and 10 are selected randomly within this interval.
The initial state is chosen to be the one with the fifth lowest energy.
The eigenvalues in the final strong sector range between 0 and 2.03, and $Y_f = $ 20 to 400 are selected.
The weak Hamiltonian that induces transitions between the 10 initial states to the $Y_f$ final states is a dense matrix with each element selected randomly.
The weak coupling constant is scaled so that
$G^2\rho_f$ is independent of the number of states, where $\rho_f$ is the density of states.
This allows for a well-defined persistence probability as $Y_f \to \infty$.
For this example, the elements of the weak Hamiltonian were chosen between $\pm w_f$, 
where $w_f = 1/(2 \sqrt{Y_F})$.
Figure~\ref{fig:ExpDecay} shows the emergence of the expected exponential decay as the number of available final states tends toward a continuous spectrum.
In a quantum simulation of a lattice theory, the density of states increases with $L$, and the late-time deviation from exponential decay will exhibit oscillatory behavior, as opposed to the plateaus found in this statistically averaged model.
The very early time behavior of the probability is interesting to note, and exhibits a well-known behavior, e.g., Refs.~\cite{Urbanowski_2017,Giacosa_2017}.  
It is, as expected, not falling exponentially, which sets in over time scales set by the energy spectrum of final states.

Only small lattices are practical for near-term simulation and lattice artifacts will be important to quantify. Relative to the continuum, a finite lattice spacing modifies the energy-momentum relation and introduce a momentum cut-off on the spectra.  
However, if the initial particle has a mass that is much less than the cut-off, these effects should be minimal as the energy of each final state particle is bounded above by the mass of the initial particle.
As has been shown in this appendix, working on a small lattice with its associated sparse number of final states, will lead to significant systematic errors when extracting the decay rates directly from the persistence probabilities.
Further work will be necessary to determine how to reliably estimate these errors.

\section{\texorpdfstring{$\beta$}{Beta}-Decay Circuits}
\label{app:BetaCircuits}
\begin{figure}[!ht]
    \centering
    \includegraphics[width=10cm]{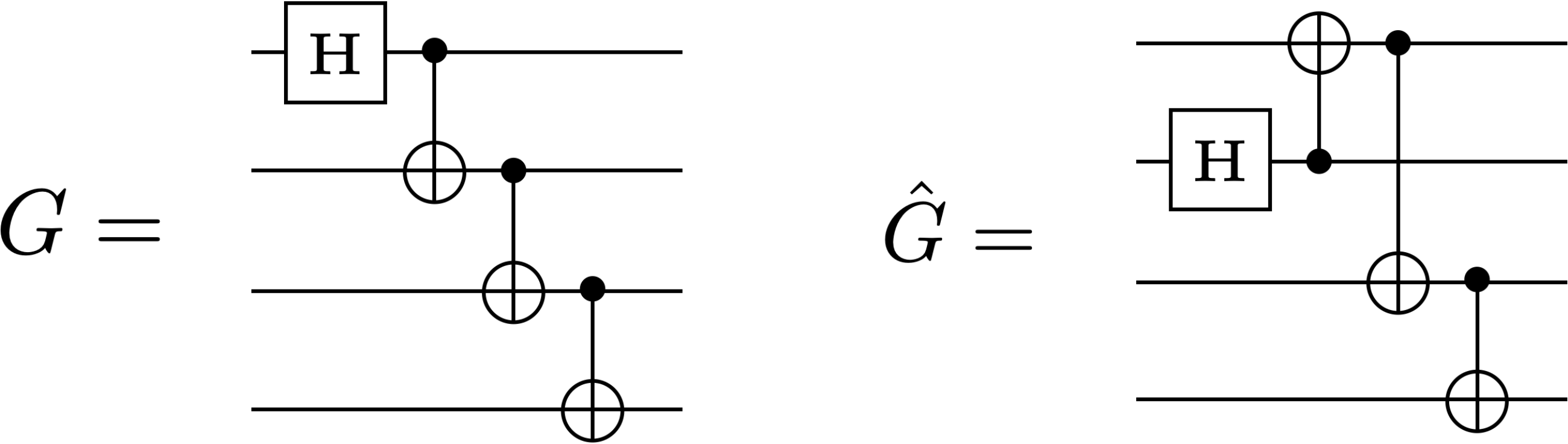}
    \caption{
    Two GHZ state preparation circuits.}
    \label{fig:GHZCirc}
\end{figure}

\begin{figure}[!ht]
    \centering
    \includegraphics[width=\columnwidth]{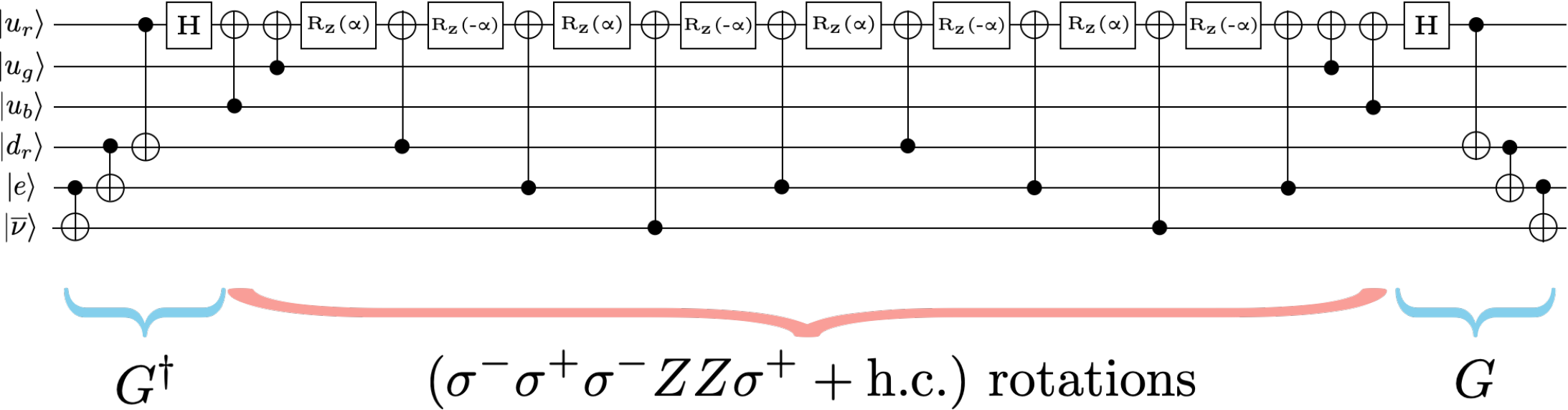}
    \caption{
    A quantum circuit that provides the time evolution associated with the 
    $\sigma^-_{\overline{\nu}} \sigma^+_e \sigma^-_{d,r} Z_{u,b} Z_{u,g} \sigma^+_{u,r}$ operator in the $\beta$-decay Hamiltonian, with 
    $\alpha = \sqrt{2} G t/8$.
    }
    \label{fig:BetaCirc}
\end{figure}

\noindent
The quantum circuits that implement the Trotterized time-evolution of the 
$\beta$-decay Hamiltonian are  similar to those
presented in Ref.~\cite{Farrell:2022wyt} 
to implement the strong-interaction dynamics,
and here the differences between the two will be highlighted.
The $\beta$-decay Hamiltonian in both the standard and tilde layouts, Eqs.~(\ref{eq:KSHamJWmap}) and~(\ref{eq:tildeBeta}), contains terms of the form
\begin{align}
    H_{\beta} \sim&\ (\sigma^- \sigma^+ \sigma^- \sigma^+ + {\rm h.c.}) + (\sigma^- \sigma^+ \sigma^+ \sigma^- + {\rm h.c.}) \nonumber \\
    =&\ \frac{1}{8}(XXXX+YYXX-YXYX+YXXY+XYYX-XYXY+XXYY+YYYY) \nonumber \\
    &+ \frac{1}{8}(XXXX + YYXX + YXYX - YXXY - XYYX + XYXY + XXYY + YYYY)
    \ ,
\end{align}
which can be diagonalized by the GHZ state-preparation circuits, $G$ and $\hat{G}$, shown in Fig.~\ref{fig:GHZCirc}. 
In the GHZ basis, it is found that
\begin{align}
    &G^{\dagger}(XXXX+YYXX-YXYX+YXXY+XYYX-XYXY+XXYY+YYYY)G \nonumber \\ 
    &= IIIZ - ZIIZ + ZZIZ - ZZZZ -IZIZ + IZZZ - IIZZ + ZIZZ 
    \ ,
\end{align}
and 
\begin{align}
    &\hat{G}^{\dagger}(XXXX + YYXX + YXYX - YXXY - XYYX + XYXY + XXYY + YYYY)\hat{G} \nonumber \\ 
    &= IIZI - ZIZI -ZZZZ+ZZZI+IZZZ-IZZI-IIZZ+ZIZZ
    \ .
\end{align}
Once diagonalized the circuit is a product of diagonal rotations, see Fig.~\ref{fig:BetaCirc} for an example of the quantum circuit that provides the time evolution associated with
$\sigma^-_{\overline{\nu}} \sigma^+_e \sigma^-_{d,r} Z_{u,b} Z_{u,g} \sigma^+_{u,r}$.
By diagonalizing with both $G$ and $\hat{G}$ and arranging terms in the Trotterization so that operators that act on the same quarks are next to each other, 
many of the CNOTs can be made to cancel.
Also, an ancilla can be used to efficiently store the parity of the string of $Z$s between the $\sigma^{\pm}$.
%

\section{Resource Estimates for Simulating \texorpdfstring{$\beta$}{Beta}-Decay Dynamics}
\label{app:LongJW}
\begin{figure}[!ht]
    \centering
    \includegraphics[width=15cm]{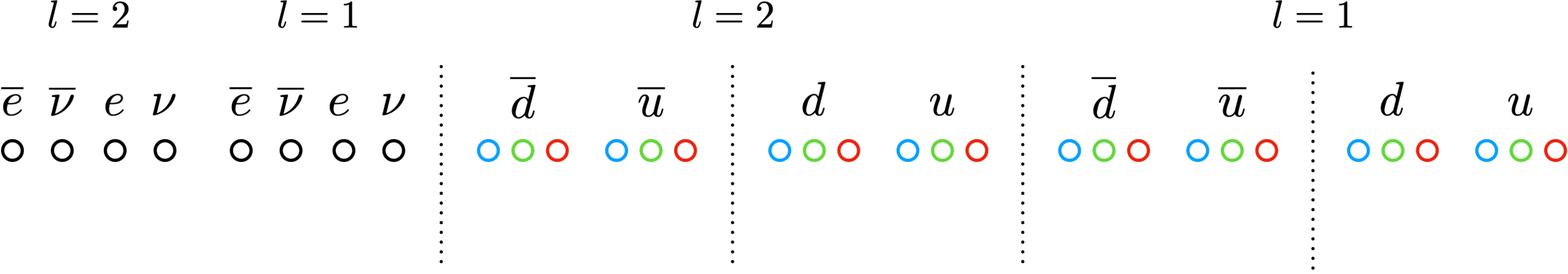}
    \caption{
    A qubit layout that is efficient for the simulation of $\beta$-decay. Shown is an example for $L=2$.}
    \label{fig:L2BetaLayout}
\end{figure}

\noindent 
For multiple lattice sites, it is inefficient to work with leptons in the tilde basis. 
This is due to the mismatch between the local four-Fermi interaction 
and the non-local tilde basis eigenstates. 
As a result, the number of terms in the $\beta$-decay component of the Hamiltonian will scale as $\mathcal{O}(L^2)$ in the tilde basis, 
as opposed to $\mathcal{O}(L)$ in the local occupation basis.
This appendix explores a layout different from the one in Fig.~\ref{fig:L1layout}, which is optimized for the simulation of $\beta$-decay on larger lattices.
To minimize the length of JW $Z$ strings, all leptons are placed at the end of the lattice, see Fig.~\ref{fig:L2BetaLayout}.
After applying the JW mapping, the $\beta$-decay operator becomes
\begin{align}
   H_{\beta} \rightarrow \frac{G}{\sqrt{2}}\sum_{l = 0}^{L-1}\sum_{c=0}^2\bigg (&\sigma^-_{l,\nub} \sigma^+_{l,e} \sigma^-_{l,d,c} Z^2 \sigma^+_{l,u,c} \: - \: \sigma^+_{l,\eb}Z^2\sigma^-_{l,\nu} \sigma^-_{l,d,c} Z^2 \sigma^+_{l,u,c} \: + \: \sigma^-_{l,\nub} \sigma^+_{l,e} \sigma^-_{l,\db,c} Z^2 \sigma^+_{l,\ub,c}  \nonumber \\ 
   &- \: \sigma^+_{l,\eb} Z^2 \sigma^-_{l,\nu} \sigma^-_{l,\db,c} Z^2 \sigma^+_{l,\ub,c} \: + \: \sigma^+_{l,e} \sigma^-_{l,\nu} \sigma^-_{l,\db,c} Z^8 \sigma^+_{l,u,c} \: - \: \sigma^+_{l,\eb} \sigma^-_{l,\nub} \sigma^-_{l,\db,c}Z^8 \sigma^+_{l,u,c} \nonumber \\
   &+ \: \sigma^+_{l,e} \sigma^-_{l,\nu} \sigma^+_{l,\ub,c} Z^2 \sigma^-_{l,d,c} \: - \: \sigma^+_{l,\eb} \sigma^-_{l,\nub} \sigma^+_{l,\ub,c} Z^2 \sigma^-_{l,d,c} \: + \: {\rm h.c.} \bigg ) \ .
   \label{eq:HBlow}
\end{align}

Using the techniques outlined in App.~\ref{app:BetaCircuits} 
to construct the relevant quantum circuits, 
the resources required per Trotter 
step of 
$H_{\beta}$ are estimated to be
\begin{align}
    R_Z  \ :& \ \ 192L \ ,\nonumber \\
   \text{Hadamard} \ :& \ \ 48L  \ ,\nonumber \\
    \text{CNOT} \ :& \ \ 436 L \ .
\end{align}
For small lattices, $L\lesssim 5$, it is expected that use of the tilde basis will be more efficient and these estimates should be taken as an upper bound.
Combining this with the resources required to time evolve with the rest of the Hamiltonian, see Ref.~\cite{Farrell:2022wyt}, the total resource requirements per Trotter step are estimated to be
\begin{align}
    R_Z  \ :& \ \ 264L^2 -54L +77 \ ,\nonumber \\
   \text{Hadamard} \ :& \ \ 48L^2 + 20L +2 \ ,\nonumber \\
    \text{CNOT} \ :& \ \ 368L^2 + 120L+74 \ .
\end{align}
It is important to note that the addition of $H_{\beta}$ does not contribute to the quadratic scaling of resources as it is a local operator. 
Recently, the capability to produce multi-qubit gates natively with similar fidelities to two-qubit gates has also been demonstrated~\cite{Katz:2022ajk,Katz:2022czu,Andrade:2021pil}.
This could lead to dramatic reductions in the resources required and, for example, the number of multi-qubit terms in the Hamiltonian scales as
\begin{equation}
    \text{Multi-qubit terms} \ : \ \ 96 L^2 -68L+22 \ .
\end{equation}
The required number of CNOTs and, for comparison, the number of multi-qubit terms in the Hamiltonian, for a selection of different lattice sizes are given in Table~\ref{tab:cnot}. 
\begin{table}[!ht]
\renewcommand{\arraystretch}{1.2}
\begin{tabularx}{0.48\textwidth}{||c | Y | Y ||}
 \hline
 $L$ & CNOTS & Multi-Qubit Terms \\
 \hline\hline
 5 & 9874 & 2082 \\
 \hline
 10 & 38,074 & 8942\\
 \hline
 50 & 926,074 & 236,622\\
 \hline
 100 & 3,692,074 & 953,222\\
 \hline
\end{tabularx}
\renewcommand{\arraystretch}{1}
\caption{The CNOT-gate requirements to perform one Trotter step of time evolution 
of $\beta$-decay for a selection of lattice sizes. For comparison, the number of multi-qubit terms in the Hamiltonian is also given.}
\label{tab:cnot}
\end{table}
Note that these estimates do not include the resources required to prepare the initial state.

\section{Technical Details on the Quantinuum {\tt H1-1} Quantum Computer}
\label{app:H1specs}
\noindent 
For completeness, this appendix contains a brief description of Quantinuum's {\tt H1-1} 20 
trapped ion quantum computer (more details can be found in~\cite{h1-1}). 
The {\tt H1-1} system uses the System Model {\tt H1} design, 
where unitary operations act on a single line of ${}^{172}$Y${}^+$ ions induced by lasers. 
The qubits are defined as the two hyperfine clock states in the ${}^2S_{1/2}$ ground state of ${}^{172}$Y${}^+$. 
Since the physical position of the ions can be modified, 
it is possible to apply two-qubit gates to any pair of qubits, 
endowing the device with all-to-all connectivity. 
Moreover, there are five different physical regions where these gates can be applied in parallel. Although we did not use this feature, it is also possible to perform a mid-circuit measurement of a qubit, i.e., initialize it and reuse it (if necessary).

The native gate set for {\tt H1-1} is the following,
\begin{equation}
    U_{1q}(\theta,\phi)=e^{-i\frac{\theta}{2}[\cos(\phi) X+\sin(\phi) Y]}\ , \quad R_Z(\lambda)=e^{-i\frac{\lambda}{2}Z}\ , \quad ZZ=e^{-i\frac{\pi}{4}ZZ} \ ,
\end{equation}
where $\theta$ in $U_{1q}(\theta,\phi)$
can only take the values $\{\frac{\pi}{2},\pi\}$, 
and arbitrary values of $\theta$ can be obtained by combining several single-qubit gates, $\tilde{U}_{1q}(\theta,\phi)=U_{1q}(\frac{\pi}{2},\phi+\frac{\pi}{2}) . R_Z(\theta) . U_{1q}(\frac{\pi}{2},\phi-\frac{\pi}{2})$. 
Translations between the gates used in the circuits shown in the main text and appendices to the native ones are performed automatically by {\tt pytket}~\cite{sivarajah2020t}.
The infidelity of the single- and two-qubit gates, as well as the error of the SPAM operations, are shown in Table~\ref{tab:H1err}.
\begin{table}[!ht]
\renewcommand{\arraystretch}{1.2}
\begin{tabularx}{0.48\textwidth}{||r | Y | Y | Y ||}
 \hline
  & Min & Average & Max \\
 \hline\hline
 Single-qubit infidelity & $2\times 10^{-5}$ & $5\times 10^{-5}$ & $3\times 10^{-4}$ \\
 \hline
 Two-qubit infidelity & $2\times 10^{-3}$ & $3\times 10^{-3}$ & $5\times 10^{-3}$ \\
 \hline
 SPAM error & $2\times 10^{-3}$ & $3\times 10^{-3}$ & $5\times 10^{-3}$ \\
 \hline
\end{tabularx}
\renewcommand{\arraystretch}{1}
\caption{
Errors on the single-qubit, two-qubit and SPAM operations, with their minimum, average and maximum values.
}
\label{tab:H1err}
\end{table}
%

\section{Time Evolution Under the Full \texorpdfstring{$\beta$}{Beta}-Decay Operator}
\label{app:betaFull}
\begin{figure}[!ht]
    \centering
    \includegraphics[width=12cm]{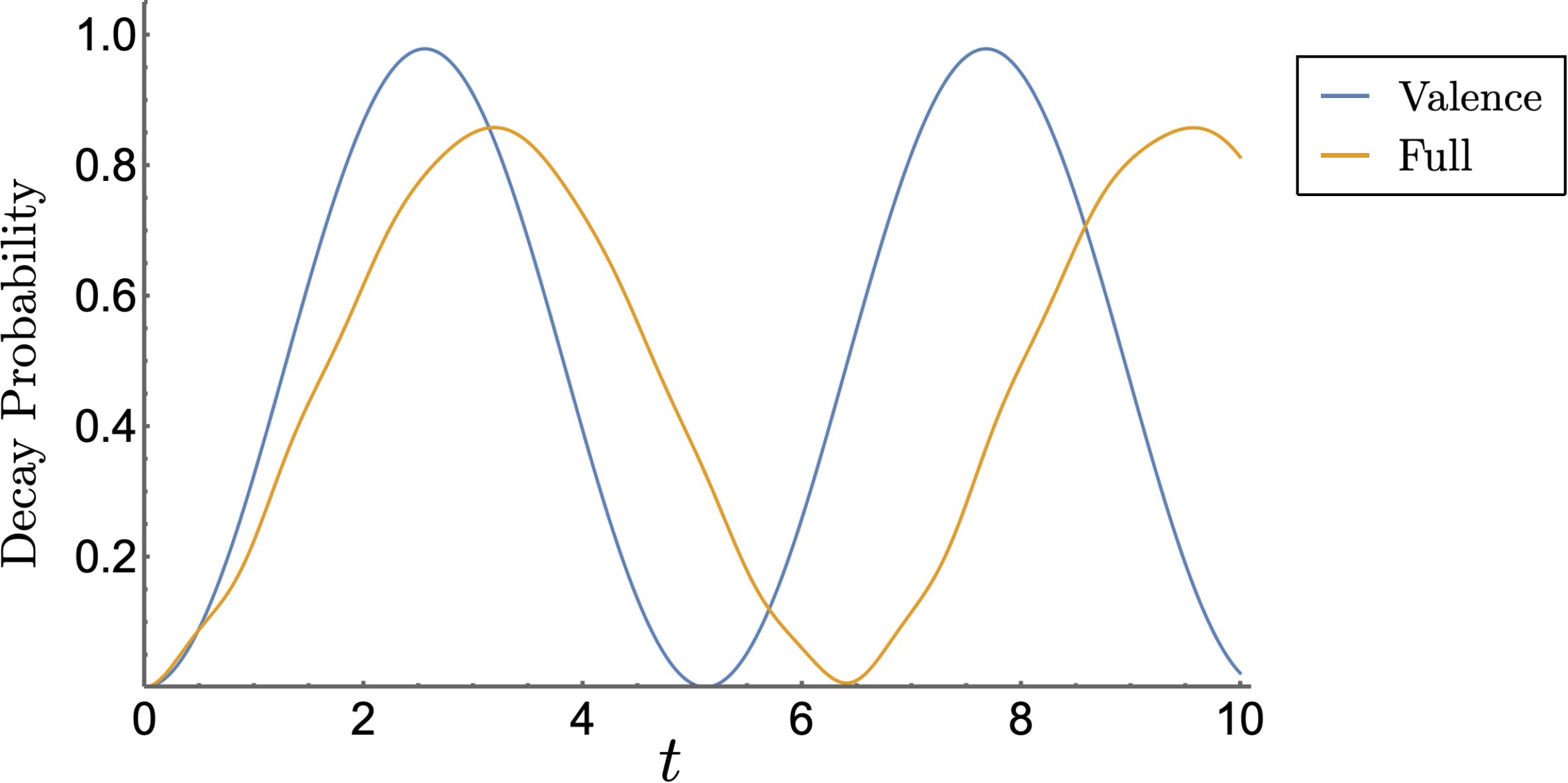}
    \caption{The probability of $\beta$-decay using both the approximate $\beta$-decay operator which only acts on valence quarks (blue) and the full operator (orange).}
    \label{fig:valfull}
\end{figure}

\noindent
The simulations performed in Sec.~\ref{sec:BetaSim} kept only the terms in the $\beta$-decay Hamiltonian which act on valence quarks, see Eq.~(\ref{eq:tildeBetaRed}). 
This appendix examines how well this valence quark $\beta$-decay operator approximates the full operator, Eq.~(\ref{eq:tildeBeta}), for the parameters used in the main text.
Shown in Fig.~\ref{fig:valfull} is the decay probability when evolved with both the approximate and full operator as calculated through exact diagonalization of the Hamiltonian.
The full $\beta$-decay operator has multiple terms that can interfere leading to a more jagged decay probability.
The simulations ran on {\tt H1-1} only went out to $t=2.5$ where the error of the approximate operator is $\sim 20\%$.

\bibliography{bibi,biblioKRS}
\end{document}